\documentclass[fleqn,usenatbib]{mnras}

\usepackage{newtxtext,newtxmath}
   
\usepackage[T1]{fontenc}
\usepackage{ae,aecompl}
\usepackage{graphicx}   % Including figure files
\usepackage{amsmath}    % Advanced maths commands
\usepackage{amssymb}    % Extra maths symbols
\usepackage{stfloats} % Add to preamble
\usepackage{makecell}
\usepackage{xcolor}
\usepackage[normalem]{ulem}

\title[FRB search of elliptical galaxies]{A targeted search for fast radio bursts from magnetars in elliptical galaxies}

\author[S. Paine et al.]{S. Paine,$^{1,2}$\thanks{E-mail: sp00048@mix.wvu.edu (SP)}
D. R. Lorimer,$^{1,2,3}$
S. Sirota,$^{1,2}$
G. M. Doskoch,$^{1,2}$
S. Tabassum,$^{1,2}$
M. Flanagan,$^{1}$
J. Lituchy,$^{4}$
\newauthor
A. Stone,$^{1,2}$
J. W. Kania,$^{1,2,5}$ 
M. Bhardwaj,$^{6,1}$ 
S. Mehta,$^{6}$
M. A. McLaughlin,$^{1,2}$
B. Kharel,$^{1,2}$
D. Agarwal$^{1,2}$
\newauthor
and
M. P. Surnis$^{7}$
\\
$^{1}$Department of Physics and Astronomy, West Virginia University, Morgantown, WV 26506-6315, USA\\
$^{2}$Center for Gravitational Waves and Cosmology, Chestnut Ridge Building, Morgantown, WV 26505-6315, USA\\
$^{3}$National Centre for Radio Astrophysics, Tata Institute of Fundamental Research, Pune, 410007, India\\
$^{4}$Morgantown High School, 109 Wilson Avenue, Morgantown, WV 26501, USA\\
$^{5}$Research IT, University of Manchester, Manchester, M13 9PY, UK\\
$^{6}$Department of Space, Planetary \& Astronomical Sciences \& Engineering, IIT Kanpur, Kanpur, UP 208016, India\\
$^{7}$Department of Physics, IISER Bhopal, Bhauri Bypass Road, Bhopal, 462066, India
}

\date{Accepted 2026 September 18. Received 2026 September 17; in original form 2026 September 3}

\pubyear{}

\begin{document}
\label{firstpage}
\pagerange{\pageref{firstpage}--\pageref{lastpage}}
\maketitle

\begin{abstract}
Motivated by predictions that globular clusters associated with nearby giant elliptical galaxies are prime environments for detecting fast radio bursts (FRBs) from magnetars, we report on a targeted high-sensitivity search across five local massive elliptical galaxies. We utilized the GREENBURST single-pulse search pipeline and the VEGAS backend on the Green Bank Telescope (GBT) to conduct an L-band (1.4 GHz) survey for FRBs in each target environment. We discovered FRB 20240817A in the direction of M49 (NGC 4472) which has a pulse width of 15 ms, a signal-to-noise ratio of 13, and a dispersion measure (DM) of $983~\rm{cm}^{-3}$ pc. Although a Bayesian spatial association framework initially flags M49 as the host, carefully accounting for the foreground electron column contributions from the Milky Way and both the Virgo intracluster medium and M49's hot gaseous halo ($\sim300-400~\rm{cm}^{-3}$ pc combined) strongly disfavours a physical connection. We conclude that FRB 20240817A is a background event. We also identified a repeating cluster of three low-significance pulses from the direction of M49 at a physically plausible DM of $150-160~\rm{cm}^{-3}$ pc. Accounting for the fraction of each target's extended globular cluster distribution observed, the absence of unambiguous, host-localized bursts allows us to place a joint upper limit on the intrinsic burst rate of a single active source within these systems of $R_0 < 0.028~\rm{hr}^{-1}$ (and an integrated rate limit of $<1.0$~hr$^{-1}$ for M49), lowering previous upper limits and suggesting that magnetar formation within dynamical clusters is less efficient than previously assumed.
\end{abstract}


\section{Introduction}

With the current sample of fast radio bursts (FRBs) now in excess of 4500 sources, it is  well established that FRBs form a cosmological population \citep[see, e.g.,][]{Petroff2019,Petroff2022}. Though most of these sources have only been observed once, over 100 known FRBs have been seen to repeat. Thanks to the emergence of FRB experiments on the Australian Square Kilometre Array Pathfinder (ASKAP), the Canadian Hydrogen Intensity Mapping Experiment (CHIME), the Deep Synoptic Array (DSA), the Five Hundred Metre Aperture Spherical Telescope (FAST), and the MeerKAT telescope, 119 FRBs have been definitively associated with a host galaxy \citep[see, e.g.,][]{Qiu2023,CHIME2018,Kocz2019,2024ApJ...967...29L}. To date, associations have been made with galaxies out to redshift $z=2.1$ \citep{2025arXiv250801648C}. This has enabled the detection and characterization of the ``missing baryons'' within the cosmic web \citep{Macquart2020,Connor2025} and a census of a variety of host galaxy environments with a range of star formation histories. 

While these localizations have led to a better understanding of the different environments that produce FRBs, there is still no consensus on the progenitors of FRBs. One of the most popular hypotheses is that FRBs arise from giant flares in the magnetospheres of magnetars, highly magnetized neutron stars whose energy source is dominated by their magnetic field  \citep{1992ApJ...392L...9D}. The strongest evidence for this connection is the FRB-like burst from Galactic magnetar SGR 1935+2154 \citep{Bochenek2020,CHIME2020}. However, several localizations challenge an exclusive magnetar origin. Notably, some repeating and non-repeating FRBs have been tracked to old, quiescent stellar populations such as globular clusters (GCs) and massive elliptical galaxies where young magnetars are not expected to form \citep{Kirsten2022,2025ApJ...979L..22E}. A larger sample of FRBs with confirmed locations would help shed light on their progenitors.

While wide-field arrays optimized to find and localize FRBs are important, high sensitivity instruments like FAST and the Robert C. Byrd Green Bank Telescope (GBT) also play an important role in FRB science. On the GBT, the GREENBURST experiment has been running since mid-2019 \citep{Surnis2019,Agarwal2020}. GREENBURST is a commensal, real-time FRB detector, running whenever the GBT is in operation, monitoring its field of view for transient radio bursts from pulsars, FRBs and other sources. So far, GREENBURST has detected more than 16,000 single pulses from 49 known pulsars \citep{Kania2026}. Three previously known repeating FRBs have also been detected as well as a new source of unknown origin and the 2.2~s pulsar J0039+5407.

The discovery of repeating FRB~20200120E \citep{Bhardwaj2021} in an M81 GC, where star formation from core-collapse supernova explosions has long since ceased, highlights the diversity of FRB host environments. While the dominant formation channel for this FRB is unclear, one promising idea is that young, FRB-producing neutron stars are produced in white dwarf mergers \citep{Lu2022} which are abundant in GC systems. This scenario is also attractive in that it could explain the long-standing puzzle of the population of young pulsars in GCs \citep{2023MNRAS.525L..22K}. Using the available constraints from the M81 discovery and follow-up observations, \citet{Kremer2023} showed that high-sensitivity facilities like the GBT have excellent prospects for uncovering more FRBs in other nearby extragalactic GCs. In addition, a recent discovery of a bright burst from FRB~20200120E with a 40-m class telescope \citep{Zhang2023} highlights the likelihood of finding similar events from sources at comparable distances.

In the summer of 2024, we began using GREENBURST to carry out a targeted search for FRBs, starting with some of the nearest galaxies. Our survey is attacking the FRB localization problem from the opposite direction of typical surveys. Rather than blindly searching for FRBs, we intend to find FRBs from particular locations. By targeting  nearby galaxies where lower luminosity sources are more likely to be detectable, we expect to place significant constraints on the poorly known FRB luminosity function, \citep[see, e.g.,][]{Fialkov2018}. Now that it is established that FRBs arise from a wide variety of environments, we anticipate that targeted surveys like this one provide complementary coverage to blind surveys probing the FRB population. Other recent targeted searches for FRBs in nearby galaxies have been carried out by \citet{NorthernCross} and \citet{2024MNRAS.534.3377C}. Both these works monitored a sample of nearby galaxies for radio transients. Each group found a background FRB, but neither found an FRB from their chosen galaxies.

The rest of this paper is written as follows. In Section \ref{sources}, we compile the sources targeted for this survey and their expected detection rates. In Section \ref{observations}, we discuss our observations and data analysis pipeline. In Section \ref{results}, we present our results, which are discussed in detail in Section \ref{discussion}. Finally, in Section \ref{conclusions}, we draw our main conclusions and make suggestions for future work.

\begin{table*}
\centering
\begin{tabular}{lcccccccccc}
\hline
{Galaxy} & {Distance} & {R.A.} & {Decl.} & {Angular size} & {$f_{\text{beam}}$} & {Predicted rate} & {$t_{\rm obs}$} & {$R_{\rm obs}$} & {$R_{\rm int}$} \\
 & { (Mpc)} & {(hh:mm:ss.s)} & {(dd:mm:ss)} & {} & {} & {(hr$^{-1}$)} & {(hr)} & {(hr$^{-1}$)} & {(hr$^{-1}$)}\\ \hline
NGC 4374 (M84) & 17 & 12:25:03.7 & +12:53:13 & $6.5^{\prime}\times5.6^{\prime}$ & 0.95 & 0.26 (0.01--0.76) & 4.0 & $<0.25$ & $<0.9$ \\ 
NGC 4472 (M49) & 17 & 12:29:46.7 & +08:00:02 & $10.2^{\prime}\times8.3^{\prime}$ & 0.33 & 0.05 (0.003--0.15) & 9.3 & $<0.11$ & $<1.0$ \\ 
NGC 4486 (M87) & 16 & 12:30:49.4 & +12:23:28 & $8.3^{\prime}\times6.6^{\prime}$ & 0.22 & 0.03 (0.002--0.09) & 2.1 & $<0.48$ & $<2.6$ \\  
NGC 4594 (M104)& \,\,\,\,\,\,\,\,9.4 & 12:39:59.4 & --11:37:23 & $8.7^{\prime}\times3.6^{\prime}$ & 0.45 & 0.15 (0.01--0.46) & 5.6 & $<0.18$ & $<0.3$ \\ 
NGC 4649 (M60) & 17 & 12:43:39.6 & +11:33:09 & $7.4^{\prime}\times6.0^{\prime}$ & 0.32 & 0.16 (0.01--0.48) & 5.9 & $<0.17$ & $<0.8$ \\
\hline
\end{tabular}
\caption{Properties of targeted galaxies. From left to right, we list the galaxy name, distance, celestial coordinates (J2000 epoch), optical angular size (semi-major and semi-minor axes at 25 B-magnitudes), the beam-capture fraction ($f_{\text{beam}}$) evaluated at 1.42 GHz, the beam-weighted predicted detection rate from \citet{Kremer2023} as described in Section~\ref{sources}, total observing time ($t_{\rm obs}$), simple rate upper limits ($R_{\rm obs}$, the reciprocal of the observing time), and the 95\% confidence upper limit on the rate ($R_{\rm int}$) obtained from the likelihood analysis after marginalizing over the power-law energy index, $\alpha$, described in Section \ref{sec:likelilhood}.}
\label{tab:rates}
\end{table*}

\section{Source selection and expected burst rates}\label{sources}

\citet{Kremer2023} estimate $\sim$60 sources with similar (or greater) fluences to FRB~20200120E within 20~Mpc. For those galaxies in the sample in which at least one source is expected, provided that the system temperature is not dominated by the galaxy itself, the inferred FRB burst rates for an experiment like GREENBURST are only an order of magnitude smaller than those from MeerKAT \citep[see Table 2 in][]{Kremer2023}. We list in Table \ref{tab:rates} five candidate target galaxies for GREENBURST for which the predicted event rate is highest: NGC~4594, NGC~4472, NGC~4486, NGC~4649, and NGC~4374.  

For the GBT Gregorian focus, the full width at half maximum of the primary beam is
\begin{equation}
\theta_{\text{FWHM}} \approx \frac{12.6}{f_{\text{GHz}}} \text{ arcmin},
\end{equation}
where $f_{\text{GHz}}$ is the observing frequency in GHz. Across our observing band, this yields a nominal beamwidth ranging from $\approx 11^{\prime}$ at the lower band edge ($1.15\text{ GHz}$) down to $\approx 7.3^{\prime}$ at the upper edge ($1.73\text{ GHz}$), with a value of $\approx 8.9^{\prime}$ at the central neutral hydrogen line frequency ($1.42\text{ GHz}$). While the optical angular sizes of our target galaxies range up to $\sim 10^{\prime}$ (such as for M49 and M104), single central pointings are utilized for each target. To quantify population losses in the outer halos beyond the primary beam, as detailed in Appendix \ref{app:fbeam}, we convolve each galaxy's projected GC spatial profile with the GBT beam response function to determine a beam-capture fraction, $f_{\rm beam}$, for each target galaxy. The resulting  $f_{\text{beam}}$ values range from $\approx 0.22$ for the expansive halo of M87 up to $\approx 0.95$ for more compact systems like M84, and are explicitly incorporated into the predicted rates for each target given in Table~\ref{tab:rates} and detailed next.

We determined the expected rates for this survey following the procedure given in \citet{Kremer2023}. Starting with the burst rate from a single source which we will call $r$, following \citet{Kremer2023}, we can calculate $r$ in terms of pulse energy ($E$), where the number of bursts as a function of energy ${\rm d}N/{\rm d}E \propto E^{\alpha}$ for some power-law index $\alpha$. For a survey with a fluence threshold $F_{\nu, \text{th}}$, observing a source at distance $D$, \citet{Kremer2023} find
\begin{equation}\label{eq:krrates}
    r \approx R_0 \, \left( \frac{F_{\nu,\text{th}}}{5~\text{Jy ms}} \right)^{\alpha + 1} \left( \frac{D}{3.6~\text{Mpc}} \right)^{2(\alpha + 1)},
\end{equation}
where this formalism is scaled by the fluence threshold of CHIME/FRB and the distance to M81 for FRB~20200120E \citep{Bhardwaj2021}. Here, the leading constant $R_0 \approx 0.07$~hr$^{-1}$ as determined by \citet{Kremer2023} from the observation of the single source in M81. For our observations,
\begin{equation}\label{eq:fluencethresh}
    F_{\nu,\text{th}} = \frac{{\rm S/N}~T}{G}\sqrt{\frac{W}{B}} = 0.37 \, \text{Jy ms},
\end{equation}
where S/N is the desired signal-to-noise ratio for a detection, in our case 10, $T = 20$ K, $G = 2$ K~Jy$^{-1}$, and $B = 960$~MHz are the system temperature, gain, and bandwidth, respectively, determined from GBT documentation, and $W$ is the pulse width, for which we use 0.011 s, the average value of FRB pulse widths from CHIME/FRB Catalog 2 \citep{CHIMECat2}. We note that 960~MHz is an ambitious usable bandwidth. A more conservative fluence calculation is 0.47 Jy~ms, calculated with a bandwidth of 500~MHz. Since Equation~\ref{eq:krrates} calculates the rate for a single FRB source, we then multiply by the effective number of sources captured within the primary beam, $N_{\text{eff}} = f_{\text{beam}} N$ (incorporating the beam-capture fractions in Table~\ref{tab:rates}). For example, in M49 we use $\alpha = -2.4$, $D = 17.03\text{ Mpc}$, $N = 7.7_{-7.3}^{+15.4}$ (with errors at the 90\% confidence level), and $f_{\text{beam}} = 0.33$. In Table~\ref{tab:rates} we quote these beam-weighted predicted rates $R = N_{\rm eff} r$ as the nominal values along with their lower and upper limits in parentheses. The calculated rates for M49 are $0.05\text{ (0.003--0.15)~hr}^{-1}$ after accounting for beam losses. Rate estimates for all galaxies, incorporating their respective $f_{\text{beam}}$ values, are listed in Table~\ref{tab:rates}. In addition to the time observed in the course of this survey, our total GREENBURST time on source for these galaxies is 4.29 hours for M84, 10.47 hours for M49, 4.98 hours for M87, 5.71 hours for M104, and 6.2 hours for M60. The vast majority of time on source for each of these galaxies is from this survey.

\section{Observations and data analysis}\label{observations}

Our observations were carried out using two methods, both of which used the L-band (1.4~GHz) receiver on the GBT. From there, the signal is split into two paths. One of these paths used the VErsatile GBT Astronomical Spectrometer \citep[VEGAS;][]{2015ursi.confE...4P} setup for pulsars, where data are saved with full Stokes parameters covering a 800~MHz band with 4096 channels centred at 1440~MHz and a time resolution of 81.92$\mu$s with 8-bit precision. The other signal path is passed to GREENBURST, which is described in detail elsewhere \citep{Surnis2019,Agarwal2020,Kania2026} and summarized below. GREENBURST data have a coarser time resolution (256 $\mu$s), but we receive processed data within $\sim$20 minutes of an observation, while the VEGAS data are processed offline. GREENBURST samples 4096 channels covering 960~MHz, centred at 1440~MHz sampled at 256~$\mu$s with 8-bit precision. 

\begin{figure}
    \centering
    \includegraphics[width=0.9\columnwidth]{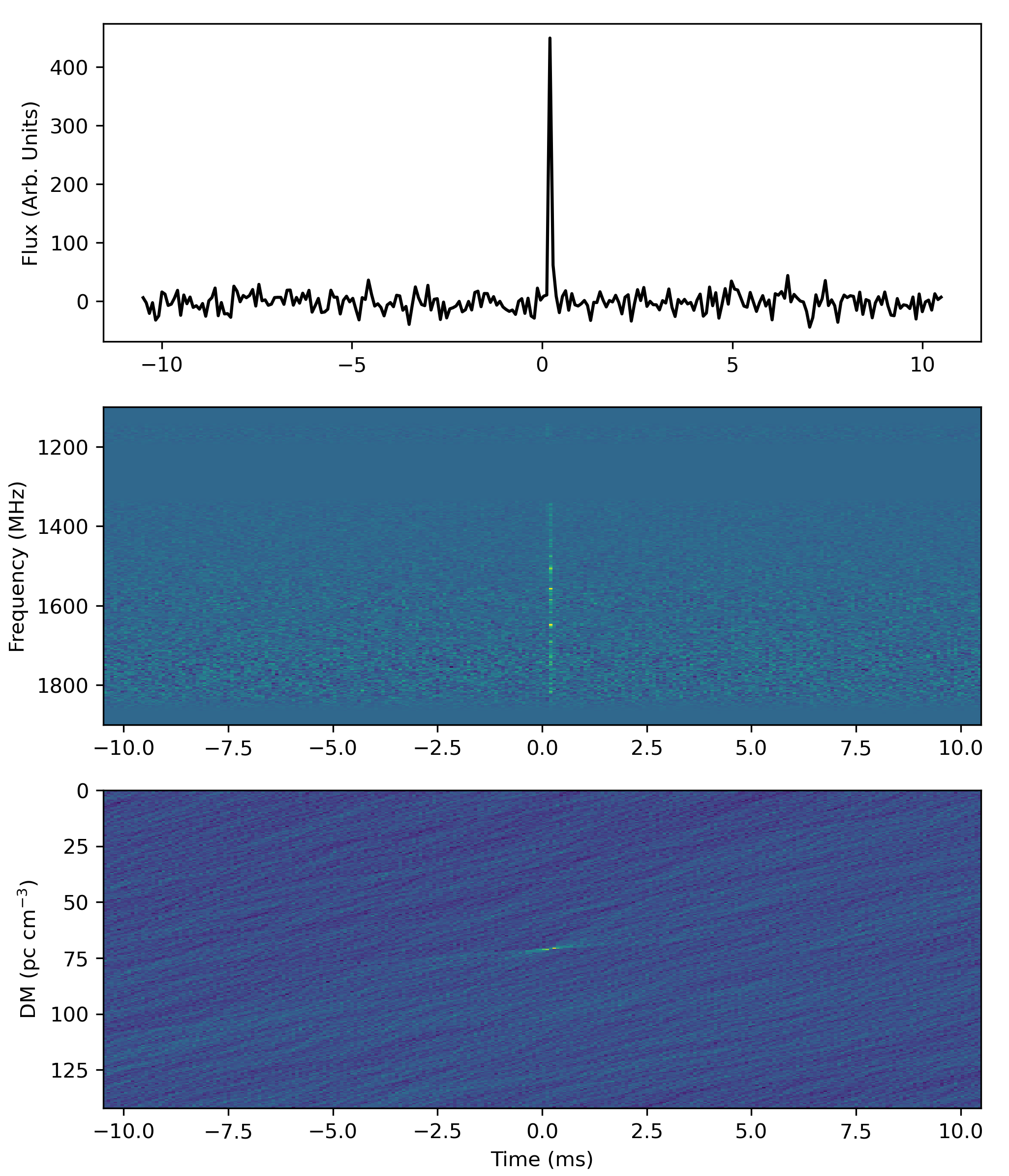}
    \caption{A single pulse from PSR~B1937+21, an excellent test source for GREENBURST, observed frequently during this project. The top panel shows dedispersed intensity versus time, the middle panel shows intensity versus dedispersed frequency and time, and the bottom panel shows intensity versus DM and time. This particular event has a S/N of 18.4.}\label{fig:expulse}
\end{figure}

The GREENBURST pipeline is designed to find astrophysical bursts of millisecond width and consists of three major parts: (i) radio frequency interference (RFI) excision; (ii) burst detection; (iii) burst categorization. RFI excision is accomplished using the Just-in-time Elimination of Spurious Signals (JESS) software, which computes the kurtosis and skew of the incoming data to determine whether signals are RFI or not \citep[for further details, see][]{Kania2026J,Kania2026}. Burst detection is accomplished with HEIMDALL\footnote{https://sourceforge.net/projects/heimdall-astro}, a package that incoherently dedisperses time-series data, followed by a single pulse search on each time series \citep{Barsdell2012}, which efficiently searches for boxcar pulses with widths in the range 256~$\mu$s to 32.8~ms with dispersion measures (DMs) ranging between 10 and 10,000 cm$^{-3}$~pc. HEIMDALL uses a S/N threshold of six. The final piece of the GREENBURST pipeline is the Fast Extragalactic Transient Candidate Hunter \citep[FETCH;][]{Agarwal2019}, a machine learning algorithm that ranks each candidate's likelihood of being an astrophysical burst \citep{Agarwal2019}. FETCH uses convolutional neural networks to analyse images of the form shown in Figure~\ref{fig:expulse}. We used FETCH's default training set, Model A, the best performing and most well-understood FETCH model, to flag promising candidates during the realtime processing stage. FETCH does not make any cutoffs based on metadata, basing its determinations of likelihood only on the visual characteristics of the plot. Metadata for all candidates, including celestial coordinates, arrival time, DM, pulse width and signal-to-noise ratio (S/N), were saved to disk.

The VEGAS data were processed using an offline version of the same pipeline used for the GREENBURST data. Since FETCH is known \citep[see, e.g.,][]{Nimmo2023} to miss weak events (S/N~$<10$), we looked over every possible candidate, marked likely by FETCH or not. We henceforth refer to the candidates FETCH identified as likely FRBs as ``FETCH-positive'' and other events as ``FETCH-negative''. In addition to searching for low S/N events, this additional layer of scrutiny through the FETCH-negative candidates was also used in recognition that FETCH is trained on previously understood events, and, at the time of FETCH's creation, several novel types of astrophysical radio bursts had not been found or understood \citep[e.g., long-period radio transients;][]{Houben2025}. As part of a concerted effort to avoid missing new phenomena, or faint FRBs not categorized by FETCH, each candidate from the above pipelines was inspected by eye by multiple authors of this paper.

Figure~\ref{fig:expulse} illustrates the diagnostic traits used during visual inspection. The top panel shows a pulse centred at zero that rises clearly above the noise floor. The middle panel displays the dedispersed frequency-versus-time plot, where a perfectly vertical signal indicates correct dedispersion. Any horizontal components are typically RFI. Finally, the bottom panel shows the DM peak, which should exhibit a ``bowtie'' falloff on both sides. While the example in Figure~\ref{fig:expulse} displays all of these ideal characteristics, most candidates did not conform perfectly to expectations. Consequently, we adopted a ranking scheme to flag candidates exhibiting obvious RFI features while prioritizing those with FRB-like traits for further analysis.

\begin{table}
    \centering
    \begin{tabular}{lcccc} 
 {Galaxy} & {MJD} & {Date} & {Start time} & {Duration} \\ 
  &  & {(MM/DD)} & {(UTC)} & {(h)} \\
 \hline
 M60 & 60536 & 08/14 & 18:47:15 & 2.0 \\
 M87 & 60536 & 08/14 & 18:53:48 & 2.0 \\
 M104 & 60536 & 08/14 & 21:14:25 & 1.7 \\
 M84 & 60538 & 08/16 & 19:02:29 & 2.0 \\
 M49 & 60538 & 08/16 & 20:04:32 & 2.0 \\
 M104 & 60538 & 08/16 & 22:18:48 & 1.9 \\
 M60 & 60539 & 08/17 & 18:05:44 & 1.9 \\
 M87 & 60539 & 08/17 & 20:11:39 & 1.9 \\
 M49 & 60539 & 08/17 & 23:05:13 & 1.0 \\
 M84 & 60540 & 08/18 & 18:12:46 & 2.0 \\
 M104 & 60540 & 08/18 & 20:13:11 & 2.0 \\
 M60 & 60540 & 08/18 & 22:46:28 & 2.0 \\
 M49 & 60546 & 08/24 & 18:10:26 & 1.3 \\
 M49 & 60551 & 08/29 & 17:09:20 & 2.1 \\
 M87 & 60551 & 08/29 & 18:46:24 & 0.2 \\
 M49 & 60574 & 09/21 & 16:59:21 & 2.9 \\
 
\end{tabular}
    \caption{From left to right, we list the galaxy observed, the Modified Julian Date (truncated to the day) of the observation, the calendar date (all during the calendar year 2024), the UTC start time, and the duration of the observation.}
    \label{tab:observations}
\end{table}

\section{Results}\label{results}

Our survey yielded 73 FETCH positive and 184,582 FETCH negative candidates through the GREENBURST pipeline, and 249 FETCH positive and 202,690 FETCH negative candidates through the VEGAS pipeline. The differences in these numbers can be attributed to the narrower frequency channels and finer time resolution of VEGAS. We visually inspected each candidate, both positive and negative from both pipelines. We found many marginal candidates that displayed some, but not all, of the characteristics of astrophysical single radio pulses. The search of the entire dataset led to the discovery of FRB~20240817A, discussed in detail in Sec.~\ref{FRB20240817A}. In addition, we found 205 marginal candidates, those that display the characteristics of an FRB, but not strongly enough to be a confident detection. These results are summarized in Table~\ref{tab:repeatdms} and discussed further in Sec.~\ref{sec:sloanebursts}.

\section{Discussion}\label{discussion}

In our interpretation of the results, we first discuss the discovery of the new FRB~20240817A and its relation to M49 and other galaxies in the GBT field of view. Next, we discuss the possibility of repeating bursts as part of an examination of the low S/N candidates. Finally, we discuss the implications of our observations compared to our expectations detailed in Section \ref{sources}.

\subsection{FRB~20240817A in the direction of M49}\label{FRB20240817A}
\begin{table}
\centering
\label{tab:dm_budget}
\begin{tabular}{lcc}
Component & Background & Member \\
 & (FRB~20240817A) & (candidate) \\
\hline
Milky Way, disk & $\sim25$ & $\sim25$ \\
Milky Way, halo & $\sim30$ & $\sim30$ \\
IGM to M49 & $\sim0.2$ & $\sim0.2$ \\
Virgo ICM & $217$ & $\sim108$ \\
M49 hot halo & $\sim40$ & \ldots \\
IGM after M49 & \ldots & unknown \\
\hline
Total foreground & $\sim310$ & $\sim165$ \\
Observed DM & $983$ & $150$--$160$ \\
Residual & $\sim670$ & $\sim0$ \\
\end{tabular}
\caption{Dispersion-measure budget (in cm$^{-3}$~pc) along the line of sight to FRB~20240817A, equivalently the direction of M49. The ``Background'' column applies to FRB~20240817A viewed through the full Virgo cluster; the ``Member'' column applies to the 150--160~cm$^{-3}$~pc candidate under the hypothesis that it lies within M49, for which only the near half of the Virgo column is intercepted and the M49 halo together with any local medium contributes to the residual rather than to the foreground.}
\end{table}

\begin{figure}
    \centering
    \includegraphics[width=0.9\columnwidth]{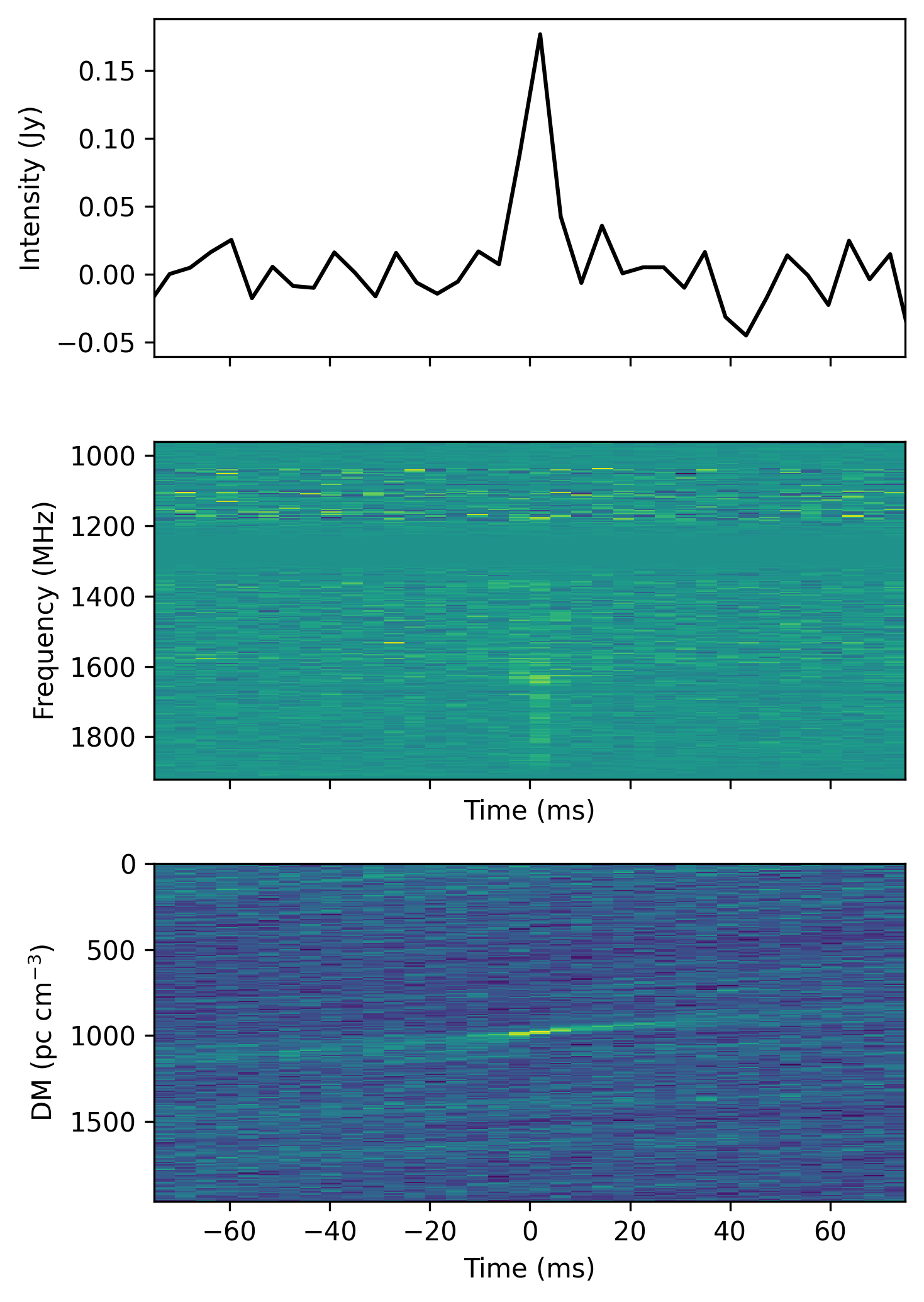}
    \caption{Detection of FRB~20240817A showing time slices of the full detection plot, with the three panels following the same pattern as those in Figure~\ref{fig:expulse}.}\label{fig:frb20240817pulse}
\end{figure}

On August 17, 2024, during an observation of Messier 49 (M49),  GREENBURST  detected a pulse of width 15~ms with a signal-to-noise ratio of 13 and a DM of 983~cm$^{-3}$~pc shown in Figure~\ref{fig:frb20240817pulse}. The pulse is consistent with all other previously known FRBs detected by GREENBURST \citep{Kania2026}, and we henceforth refer to it as FRB~20240817A. Unfortunately, during the time of detection, VEGAS data were not being recorded, and the GREENBURST data shown in Figure~\ref{fig:frb20240817pulse} is the highest resolution available for this source. Of more than 7,000 candidates with S/Ns above 13, every other candidate is obviously RFI. In the absence of RFI, the expected number of pulses can be computed from Gaussian statistics \citep[see, e.g.,][]{Cordes2003}. For the number of DM trials in our survey (1451), we find that less than one candidate with a S/N greater than 10 would be expected over the 7 year lifetime of GREENBURST, i.e.~far longer than the current survey. In addition, based on the similarities of this detection with known FRBs from GREENBURST reported by \citet{Kania2026}, we are confident in the astrophysical origin of this signal.

\begin{figure}
    \centering
    \includegraphics[width=0.9\columnwidth]{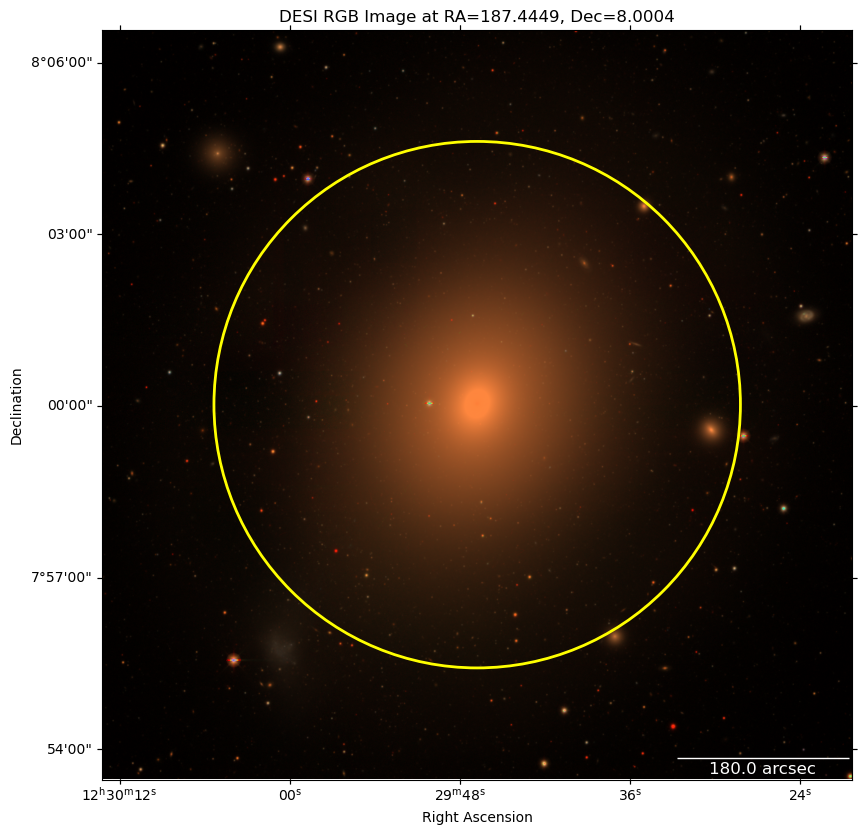}
    \caption{DESI g, r, z composite image of the search field showing the region around M49. The large galaxy at the center of the field is M49. The overlaid circle (radius = 4.6 arcmin) represents the GBT beam at L-band and encloses 228 catalogued galaxies (Table~\ref{tab:galaxyIDs}).}\label{fig:M49}
\end{figure}

In Figure~\ref{fig:M49}, we construct an RGB composite image of the field around M49 using DESI \citep[Dark Energy Spectroscopic Instrument;][]{Dey2019} imaging in the g, r, and z bands, mapped to the blue, green, and red channels, respectively. The localization region of FRB~20240817A contains the giant elliptical galaxy M49, making it a plausible host candidate. However, M49 resides within the Virgo cluster. As a result, a significant fraction of the observed dispersion measure (DM) may arise from ionized gas associated with the Virgo environment. Consequently, the observed DM cannot be directly interpreted as a host-galaxy contribution without accounting for the foreground electron column associated with both the Virgo intracluster medium (ICM) and the hot gaseous halo of M49.

The Virgo cluster is a dynamically young, highly structured system whose dominant gravitational potential is centred on M87, which also lies near the peak of the cluster X-ray emission \citep{1985AJ.....90.1681B,1994Natur.368..828B,2011MNRAS.414.2101U}. To estimate the cluster contribution, we adopt the electron density profile derived by \citet{Agarwal2019MNRAS} from \textit{Planck} measurements of the thermal Sunyaev--Zel'dovich effect in which the electron density is given by
\begin{equation}
n_e(b,z) = 8.5\times10^{-5} \left(b^2+z^2\right)^{-0.6} \ {\rm cm}^{-3},
\end{equation}
where $b$ is the projected distance from the Virgo cluster centre, $z$ is the line-of-sight coordinate, and both quantities are in Mpc. Following \citet{Agarwal2019MNRAS}, the corresponding dispersion measure is
\begin{equation}\label{eq:dmvirgo}
{\rm DM}_{\rm Virgo} = 10^6 \int_{-2.4}^{+2.4} n_e(b,z)\,{\rm d}z,
\end{equation}
where the factor of $10^6$ converts Mpc to pc and the integral is carried out over two virial radii (2.4~Mpc). Since M49 lies approximately $1.25$~Mpc in projection from M87, the Virgo ICM contributes substantially to the total electron column even for sightlines passing directly through M49. Evaluating Equation~\ref{eq:dmvirgo} at the projected location of M49, or $b=1.25$ Mpc, yields ${\rm DM}_{\rm Virgo} \simeq 217~{\rm cm^{-3}\,pc}$.

In addition to the Virgo ICM, we consider the hot X-ray-emitting medium surrounding M49. Massive elliptical galaxies are known to host extended reservoirs of hot ($T\sim10^7$~K) gas that dominate their baryonic content at large radii \citep{2003ARA&A..41..191M}. M49 is one of the best-studied examples, with deep \textit{Chandra} and \textit{XMM-Newton} observations revealing a smooth and extended gas distribution extending to radii of at least $\sim30$~kpc \citep{Irwin1996ApJ,Kraft2011ApJ}. Following \citet{Kraft2011ApJ}, we estimate the halo contribution by integrating the deprojected electron-density profile along the line of sight. This calculation properly accounts for the geometry of the sightline through the halo rather than assuming that the projected offset is equivalent to the three-dimensional radius.

The resulting halo contribution is expected to be at most a few times $10^{1-2}$~${\rm cm^{-3}\,pc}$ for sightlines passing near the centre of M49, and decreases rapidly with increasing impact parameter. In addition to these extragalactic screens, the line of sight also crosses the Milky Way. Towards the Galactic coordinates of M49 ($l \simeq 287^{\circ}$, $b \simeq +70^{\circ}$), the disk models of \citet{yao2017}, \citet{ne2025}, and \citet{cordes2002} predict Galactic contributions of only $21$, $27$, and $29~{\rm cm^{-3}\,pc}$ respectively (i.e.\ ${\rm DM_{MW,disk}} \approx 25~{\rm cm^{-3}\,pc}$), with negligible scattering. To these estimates, we add an uncertain Galactic-halo term of $\sim30~{\rm cm^{-3}\,pc}$ that these disk models do not include. A second consideration is that the relevant Virgo column depends on where the source lies along the line of sight: a background source is seen through the entire cluster and intercepts the full column of Equation~\ref{eq:dmvirgo} ($217~{\rm cm^{-3}\,pc}$), whereas a source located within M49 samples only the near side, i.e.\ roughly half this value ($\sim108~{\rm cm^{-3}\,pc}$). Both cases are summarised in Table~\ref{tab:dm_budget}.

For FRB~20240817A the relevant total is the ``Background'' column. Summing the Galactic disk and halo, the full Virgo ICM, the hot halo of M49, and the negligible intergalactic medium (IGM) contribution gives a foreground of only $\sim310~{\rm cm^{-3}\,pc}$, far short of the observed $983~{\rm cm^{-3}\,pc}$ and leaving $\sim670~{\rm cm^{-3}\,pc}$ to be supplied by the intergalactic medium and a redshifted host. The large DM therefore cannot be produced by the Virgo environment, and M49, being an elliptical galaxy, does not host enough gas to plausibly supply the DM gap. We conclude that FRB~20240817A is a background source viewed through the cluster rather than a member of M49.

\subsection{Search for the galaxy host of FRB 20240817A}\label{sec:search}

We utilise the $4.6\arcmin$ single-beam localisation to evaluate host candidates for FRB 20240817A using the Probabilistic Association of Transients to their Hosts (PATH) framework, as described by \citet{Aggarwal2021}. A limitation of the current framework is its inability to accurately represent the localisation distribution for single-beam measurements. PATH hard-codes a two-dimensional Gaussian probability density, whereas a single-beam detection only constrains the FRB to lie within the beam footprint and does not naturally correspond to a Gaussian profile. We therefore modify the framework to approximate the localisation using a circular top-hat distribution. We identify host candidates using galaxy effective radii and apparent $r$-band magnitudes from the DESI Legacy Imaging Surveys Tractor catalogue. The single-beam localisation is centred within $0.5\arcsec$ of M49. Because M49 is exceptionally bright ($m_r$ = AB mag) and spatially extended ($R_{\mathrm{eff}} \approx 82\arcsec$), its high surface brightness obscures faint high-redshift galaxies over a large area. While the PATH default unseen prior is $P(U) = 0.1$ for deep unobstructed surveys, the local degradation in catalogue completeness due to M49 motivates us to adopt a more conservative unseen prior of $P(U) = 0.2$. The PATH analysis identifies M49 as the highest-ranked candidate host of FRB 20240817A with a posterior probability of $P(O_i|x) = 0.835$. A detailed discussion of the PATH posterior probability is presented in Appendix \ref{appendix1}. However, as discussed in Section~\ref{FRB20240817A}, our analysis of the foreground DM contribution from the Virgo environment indicates that the observed DM cannot be explained by an origin within M49. We again conclude that FRB~20240817A is associated with a background source viewed through the Virgo cluster.

\subsection{Marginal candidates in the survey}\label{sec:sloanebursts}

\begin{figure*}
    \centering
    \includegraphics[width=1.0\textwidth]{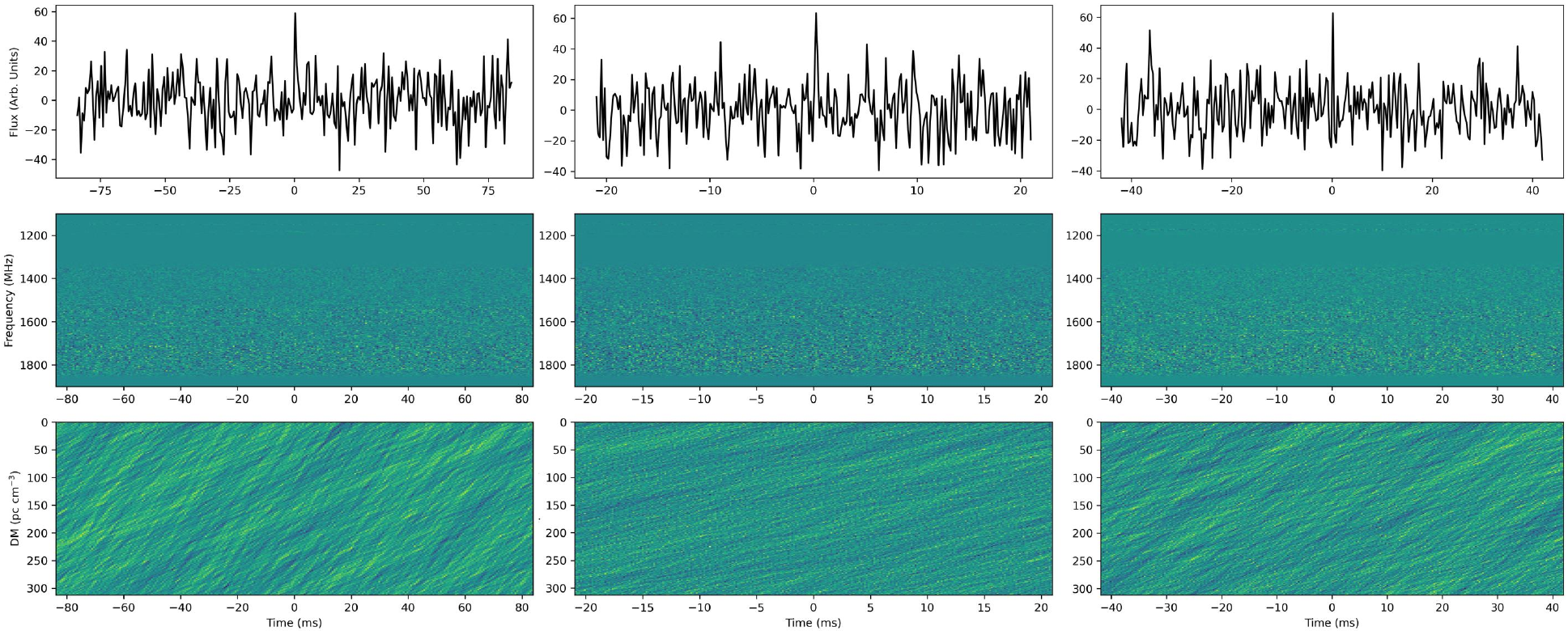}
    \caption{Diagnostic plots for the three marginal candidate bursts with DMs between $150$ and $160$~cm$^{-3}$~pc. For each burst, the subpanels display the frequency-integrated pulse profile (top row), intensity as a function of time and frequency (middle row), and intensity as a function of time and trial DM (bottom row). The peaks in the top panel are clear, faint vertical lines are seen in the middle panel, and there is a peak in DM in the third panel. However, none of these DM peaks are very strong, and they do not show the characteristic bow-tie shape. Of over 400,000 candidates, only 204 look similar to these examples. The DM range 150--160~cm$^{-3}$~pc is the only one with more than two candidates that are visually non-RFI.}\label{fig:repeater}
\end{figure*}

From human inspection of the almost 400,000 candidates mentioned in Sect.~\ref{results}, 205 were determined to be possibly astrophysical. These candidates (a representative example is in Figure~\ref{fig:repeater}) have lower brightnesses and signal-to-noise ratios than the detections shown in Figures~\ref{fig:expulse} and \ref{fig:frb20240817pulse}. The S/Ns of these candidates range from 6.5 to 8.3. The assignment of FETCH-negative is not  due to low S/N, but relies on FETCH's determination of the characteristics of the plots. Table~\ref{tab:repeatdms} lists the low-significance candidates seen across our target galaxies.

\begin{table}
    \centering
    \begin{tabular}{lcc} 
 {Galaxy} & {Number of pulses} & {DM range}\\ 
  &  & {(cm$^{-3}$~pc)} \\
 \hline
 M49 & 74 & 10--1550 \\
 M60 & 44 & 11--990 \\
 M84 & 55 & 65--1750 \\
 M87 & 23 & 47--1060 \\
 M104 & 7 & 12--330 \\
 
\end{tabular}
    \caption{From left to right, we list the galaxy observed, the number of marginal candidates found, and the DM range of the candidates.}
    \label{tab:repeatdms}
\end{table}

Despite finding many marginal candidates in this analysis, none of them are compelling enough to count as an FRB detection in the absence of any supporting evidence. As indicated in earlier single-pulse searches \citep[see, e.g.,][]{2006Natur.439..817M}, faint repeating sources can be found from an analysis of the clustering in DM. With enough bursts, even if they are faint, there might be a case for the presence of a repeating FRB based on a statistically significant DM excess in the candidates from the target galaxies. The largest number of bursts within 10 DM units is three bursts from the direction of M49 in the DM range 150--160~cm$^{-3}$~pc. As detailed in our earlier study of M82 \citep{Paine2024}, we compute a confidence level $C$ in a detection of $n$ or more real pulses in a DM range out of a sample of $N$ pulses in a search using $T$ DM trials, finding that
\begin{equation}\label{eq:confidence}
    C = \sum_{i=0}^{n-1} \frac{(N/T)^i \exp(-N/T)}{i!}.
\end{equation}
In our case, $n=3$ from a sample of $N=74$ pulses over $T=1451$ DM trials. Our measured confidence is $C=99.998\%$, corresponding to an equivalent Gaussian significance of slightly above $4\sigma$. We do not consider this significant enough to constitute a detection. However, the repetition of these three bursts opens the possibility of a repeating FRB from the direction of M49. Further detections would be necessary to claim the discovery of a repeating FRB.

The measured DM of this candidate is compatible with an origin in M49. For a source residing in the galaxy, the expected foreground is the ``Member'' column of Table~\ref{tab:dm_budget}: the Galactic disk ($\sim25$~cm$^{-3}$~pc), the Galactic halo ($\sim30$~cm$^{-3}$~pc), and the near-side Virgo ICM ($\sim108$~cm$^{-3}$~pc) together contribute $\sim165$~cm$^{-3}$~pc, essentially the whole of the observed 150--160~cm$^{-3}$~pc. Such a source would therefore add little or no dispersion of its own, exactly as expected for a globular-cluster source analogous to FRB~20200120E, whose local environment contributes negligibly to its DM \citep[e.g.][]{Bhardwaj2021,Nimmo2023}. We caution that the DM budget available for M49's halo and any local medium is consistent with zero, but sensitive both to the poorly constrained Galactic-halo term and to the unknown line-of-sight position of M49 within Virgo.

While the GBT may not be able to detect the full number of bursts from this source, other telescopes (such as FAST, CHIME/FRB, DSA, MeerKAT, and/or SKA) may be able to place tighter constraints on this putative source. No sources from any of these galaxies have reported in CHIME/FRB Catalog 2.

\subsection{Implications for FRBs from extragalactic GCs}
\label{sec:likelilhood}

This survey was motivated by the encouraging predictions made by \citet{Kremer2023} for FRBs detectable with a 100-m class radio telescope in GC systems associated with nearby giant elliptical galaxies. Based on the preceding discussion, it is most likely that GREENBURST has no credible detections of FRBs from any of our targets to date. This null result,
therefore, constrains the rate estimates from \citet{Kremer2023}. Specifically, we disfavour the higher end of the rate estimates as shown in Table \ref{tab:rates}. For example,
in M49, our upper limit of 0.11~hr$^{-1}$ implies that the putative source population for GCs associated with that galaxy lie at the lower end of the range predicted
by \citet{Kremer2023}. With the exception of M87, our observations essentially rule out the upper ranges. This implies that perhaps FRBs are less likely to arise from magnetars than previously assumed, or that magnetars are less likely to be formed in GCs than \citet{Kremer2023} supposed.

\begin{figure}
    \centering
    \includegraphics[width=0.95\columnwidth]{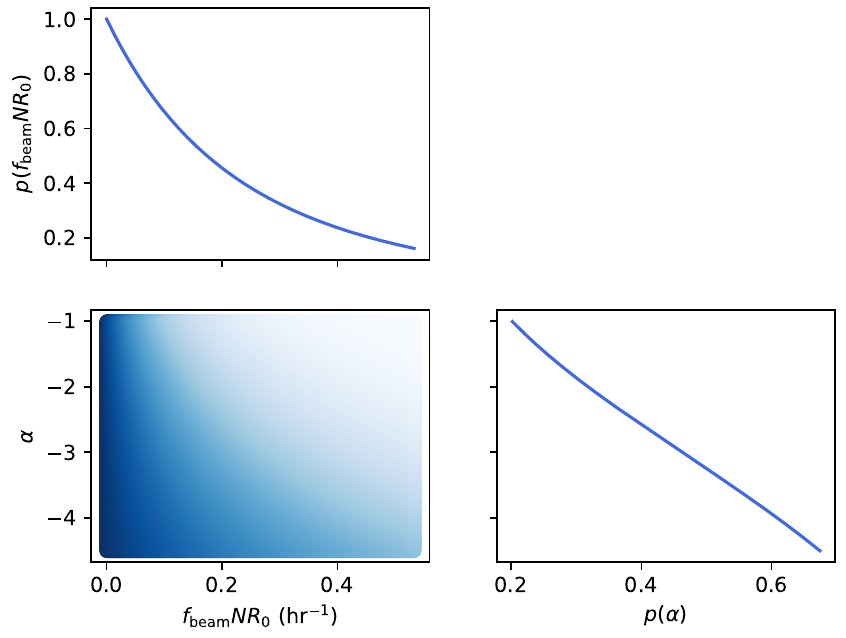}
    \caption{For M49, the likelihood of different combinations of $NR_0$, the burst rate from the entire galaxy, and $\alpha$, the power-law energy index of the sources. Darker regions of the $\alpha -NR_0$ plane indicate higher likelihood. Likelihoods are also marginalised along each axis to determine the posterior probability density function of each of the variables.}\label{fig:M49hist}
\end{figure}

Since the predicted rates listed in Table \ref{tab:rates} assume fixed values for the power-law index of pulse energies ($\alpha$) and the leading constant ($R_0$) in Equation \ref{eq:krrates}, we carry out a simple likelihood analysis in terms of $\alpha$ and $R_0$ to investigate how our results jointly constrain these parameters. 
Assuming a population of $N$ sources in the GCs of each target galaxy, of which an effective fraction $N_{\text{eff}} = f_{\text{beam}} N$ is captured within the primary beam, the probability that we have seen no bursts in some observation time $T$ is simply $P = e^{-f_{\text{beam}} Nr T}$, where $r$ is calculated using Equation~\ref{eq:krrates}. Since $N_{\text{eff}}$ and $r$ are degenerate with one another, our analysis effectively explores the parameter space defined by $\alpha$ and the product $f_{\text{beam}} N R_0$. We consider a range for $-1\le \alpha \le -4.5$,
 as suggested by multiple past works \citep{Luo2018,Nimmo2023}. The result of this analysis is shown for the case of M49 in Figure~\ref{fig:M49hist}. 
 Since we did not detect a burst from M49, $f_{\text{beam}} N R_0 = 0$ is the most likely value, with a 95\% confidence level upper limit of $f_{\text{beam}} N R_0 < 1.0~\text{hr}^{-1}$. Due to the form of Equation~\ref{eq:krrates} and our choice of fluence threshold and distance, more negative values of $\alpha$ are preferred. Integrated upper limits derived from the observations of the five target galaxies are listed in Table~\ref{tab:rates}.

\begin{figure}
    \centering
    \includegraphics[width=0.95\columnwidth]{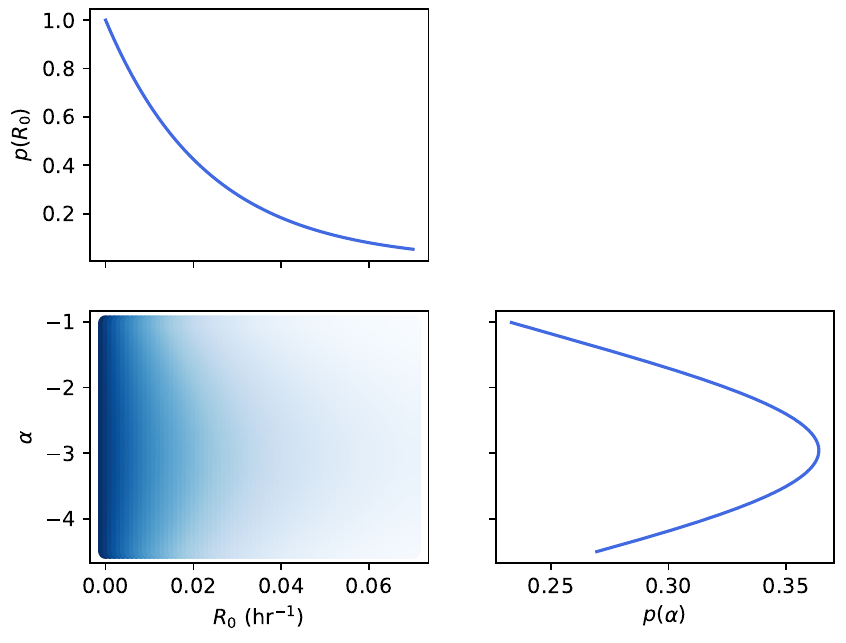}
    \caption{Likelihood of different combinations of $R_0$, the burst rate from an FRB source, and $\alpha$, their power-law energy index. As detailed in the text, this analysis assumes individual numbers of sources in the target galaxies given in \citet{Kremer2023}. Likelihoods are also marginalised along each axis to determine the posterior probability density function of each of the variables.}\label{fig:combhist}
\end{figure}

Assuming that the values of $N$ and the beam-capture fractions $f_{\text{beam}}$ found for each of the target galaxies in our sample are reasonable, we can find the likelihood of finding no sources in our survey as a function of $R_0$ and $\alpha$. By taking the product of these individual likelihoods in the $R_0 - \alpha$ plane for all five galaxies, we can attempt to constrain $R_0$ and $\alpha$ jointly. The result of this analysis is shown in Figure~\ref{fig:combhist}, which suggests that $R_0<0.028$~hr$^{-1}$ at the 95\% confidence level and that the most likely value of $\alpha$ is around $-3$.

Ultimately, these joint constraints indicate that while our non-detections do not completely falsify the magnetar progenitor model proposed by \citet{Kremer2023}, they actively narrow the permissible physical regimes for FRB production in GCs. The preference for a steeper energy distribution ($\alpha \approx -3.1$) combined with a lowered scaling constant ($R_0 < 0.028\text{ hr}^{-1}$) suggests that the highly active population archetype modeled after the M81 GC source, FRB~20200120E, may represent an observational outlier rather than the average for extragalactic cluster systems. On the other hand, FRB~20200120E is much closer than putative sources from our chosen galaxies. The distances to these galaxies range from 9.4 to 17 Mpc, in comparison to M81's 3.6~Mpc distance. Only the very brightest bursts from FRB~20200120E would be detectable at these greater distances. In fact, the detectable fraction of bursts is only around 4\%. The only three potentially detectable bursts are those from the discovery paper \citep{Bhardwaj2021}. The other 71 bursts from FRB~20200120E published to date, including the 53 from the burst storm \citep{Nimmo2023}, would not be detectable with GREENBURST at the distances of our target galaxies.

\subsection{Comparison to other work}

Upon completing this project, we became aware of a recent study by \citet{Ho2026} detailing a targeted survey of M49 for FRBs with FAST. Given its larger collecting area, FAST achieves a significantly lower fluence threshold ($16.5~\text{mJy~ms}$ vs. our $470~\text{mJy~ms}$). While covering M49's extended GC system required \citet{Ho2026} to employ a four-pointing mosaic, our GBT survey utilized a single central pointing whose primary beam captures an effective fraction $f_{\text{beam}} \approx 0.33$ of the galaxy's total globular cluster population. 

Using zero detections to establish a 95\% confidence level Poisson upper limit, we find a per-GC rate upper limit of:
\begin{equation}
    \mathcal{R}_{\mathrm{lim}} = \frac{3}{f_{\text{beam}} N_{\mathrm{GC}} \, t_{\mathrm{obs}}} \approx 1.4 \times 10^{-4}~\text{FRBs GC}^{-1}~\text{hr}^{-1},
\end{equation}
where the factor of 3 appropriately accounts for this confidence level, $N_{\mathrm{GC}} \simeq 7,000$ is the total number of GCs in M49 \citep{Kremer2023} and $t_{\mathrm{obs}} = 9.3~\text{hr}$. This compares to the limit of $4.7 \times 10^{-4}~\text{FRBs GC}^{-1}~\text{hr}^{-1}$ reported by \citet{Ho2026}. Although our GBT survey had a higher fluence threshold, our longer dwell time (\citealt{Ho2026} had $2.1~\text{hr}$ per GC) compensates for the truncated outer halo coverage, yielding a competitive rate constraint that probes the bright end of the burst energy distribution. Finally, we note that \citet{Ho2026} do not report any low-significance bursts that are consistent with those discussed in Sec.~\ref{sec:sloanebursts}.

\section{Conclusions}\label{conclusions}

Motivated by the discovery of the GC-associated FRB~20200120E in M81 \citep{Bhardwaj2021} and subsequent theoretical modeling of dynamical magnetar formation channels by \citet{Kremer2023}, we carried out a targeted search for fast radio bursts across five massive local elliptical galaxies (M49, M60, M84, M87, and M104) utilizing the GBT. While no definitive, host-associated FRB population was uncovered, this survey establishes the most stringent upper limits to date on the single-pulse rates from sources within these mature stellar environments. Specifically: (i) we place a joint upper limit on the intrinsic baseline burst rate of a single active source within these systems of $R_0 < 0.028~\text{hr}^{-1}$; (ii) for each galaxy, we place upper bounds on the beam-weighted integrated burst rate, $f_{\text{beam}} N R_0$; (iii) we discovered FRB~20240817A during the M49 observations, while a spatial probability analysis indicates that M49 is the host of FRB~20240817A, a DM budget analysis strongly disfavours a physical association, leading us to conclude that FRB~20240817A must arise from a background galaxy; (iv) we identified a marginal repeating FRB candidate at a dispersion measure of $150-160~\text{cm}^{-3}~{\rm pc}$ potentially within M49, which warrants further follow-up.

Should future monitoring reveal FRB~20240817A to be a repeating source, its utility would extend far beyond population statistics and serve as an invaluable, high-DM probe of the intervening magneto-ionic environment. 
Because the line of sight to FRB~20240817A intersects the halo of M49, continuous or polarimetric follow-up of repeat bursts could place strong constraints on the electron density, magnetic field geometry, and plasma turbulence within the circumgalactic medium of a massive elliptical galaxy, turning a chance alignment into a useful cosmological tool.

Our rate constraints effectively truncate the optimistic upper ranges predicted by \citet{Kremer2023}. The reduction of the expected burst rate highlights a gap in our understanding of either magnetar formation efficiency via dynamical channels or the true shape of the low end of the FRB luminosity function. To break the current degeneracies between true population sizes and intrinsic burst rates, and to capture the significant fractions of globular cluster populations residing outside single primary beam footprints ($f_{\text{beam}} < 1$), deeper and spatially expanded observations are needed. The detection rates for FAST predicted by \citet{Kremer2023} still remain largely untested. Moving forward with the GBT, leveraging wide-field mosaic pointings as well as the sub-millisecond time resolution of our offline VEGAS pipeline will be crucial to definitively proving or ruling out the presence of a hidden population of faint, repeating transients within these massive elliptical hosts.

\section*{Acknowledgements}

The Green Bank Observatory (GBO) is a facility of the National Science Foundation (NSF) operated under cooperative agreement by Associated Universities, Inc. We thank the GBO staff for their support of GREENBURST, particularly Ryan Lynch and Evan Smith for their assistance with the VEGAS observations. GREENBURST operations have been supported by NSF awards 1616042 and 2406570.
SP would like to thank CP for his lifelong support. DRL and MAM acknowledge support from
the Eberly family.
%%%%%%%%%%%%%%%%%%%%%%%%%%%%%%%%%%%%%%%%%%%%%%%%%%
\section*{Data Availability}

All GREENBURST triggers are available upon reasonable request. Select processed data from the pipeline for detections of known pulsars are available at https://greenburstwvu.github.io.

%%%%%%%%%%%%%%%%%%%% REFERENCES %%%%%%%%%%%%%%%


\bibliographystyle{mnras}
\bibliography{refs}

%%%%%%%%%%%%%%%%%%%%%%%%%%%%%%%%%%%%%%%%%%%%%%%%%%

%%%%%%%%%%%%%%%%% APPENDICES %%%%%%%%%%%%%%%%%%%%%

%%%%%%%%%%%%%%%%%%%%%%%%%%%%%%%%%%%%%%%%%%%%%%%%%%

\appendix
\section{Beam-capture fraction}\label{app:fbeam}

To quantify the fraction of each target galaxy's GC system captured within our single-beam observations, we compute the beam-capture fraction, $f_{\text{beam}}$, for each galaxy. We model the projected surface density profile of each galaxy's GC system, $\Sigma(\vec{\theta})$, where $\vec{\theta}$ is the angular offset from the beam centre on the sky, adopting standard spatial distributions scaled by the optical properties and effective half-number radii cataloged in previous optical surveys.

For our giant elliptical targets (NGC~4374, NGC~4472 and NGC~4649), the projected GC surface density is modeled using a projected Sérsic profile or a generalized power-law distribution \citep[see, e.g.,][]{2005PASA...22..118G}. Specifically, we adopt a projected Sérsic form for the GC number density:
\begin{equation}
    \Sigma(R) = \Sigma_0 \, \exp\left( -b_n \left[ \left(\frac{R}{R_{\text{eff}}}\right)^{1/n} - 1 \right] \right),
\end{equation}
where $R = |\vec{\theta}|$ is the projected galactocentric radius, $R_{\text{eff}}$ is the effective half-number radius of the GC system, $n$ is the Sérsic index describing profile curvature, and $b_n$ is a constant dependent on $n$. For M87 (NGC~4486), which exhibits extended halo power-law tails, we adopt a projected power-law surface density profile $\Sigma(R) \sim R^{-\beta}$ with $\beta \approx 1.5$ out to the truncation radius of the cluster system. For the lenticular galaxy NGC~4594 (M104), we adopt a composite two-component surface density profile that accounts for both its prominent spheroidal bulge and its flattened stellar disk components.

The GBT primary beam response at our central observing frequency ($1.42\text{ GHz}$) is approximated as a two-dimensional circular Gaussian with a full width at half maximum of $\theta_{\text{FWHM}} \approx 8.9^{\prime}$. The normalized beam response function 
\begin{equation}
    B(\vec{\theta}) = \frac{1}{2\pi \sigma_{\text{beam}}^2} \exp\left( -\frac{|\vec{\theta}|^2}{2\sigma_{\text{beam}}^2} \right),
\end{equation}
where $\sigma_{\text{beam}} = \theta_{\text{FWHM}} / \sqrt{8 \ln 2}$ is the beam dispersion. 

The effective beam-capture fraction is then defined as the ratio of the integrated convolved cluster distribution to the total integrated population which
can be written as
\begin{equation}
    f_{\text{beam}} = \frac{\int \Sigma(\vec{\theta}) B(\vec{\theta}) \, {\rm d}^2\theta}{\int \Sigma(\vec{\theta}) \, {\rm d}^2\theta}.
\end{equation}
This integration accounts for geometric flux and source losses in the outer galactic halos, yielding the effective source counts $N_{\text{eff}} = f_{\text{beam}} N$ utilized in our rate constraints in Table \ref{tab:rates}.

\section{PATH analysis of FRB 20240817A}
\label{appendix1}

\begin{figure*}
    \centering
    % Figure 1a (Left) - Reduced width so they fit on one line
    \includegraphics[width=0.45\textwidth]{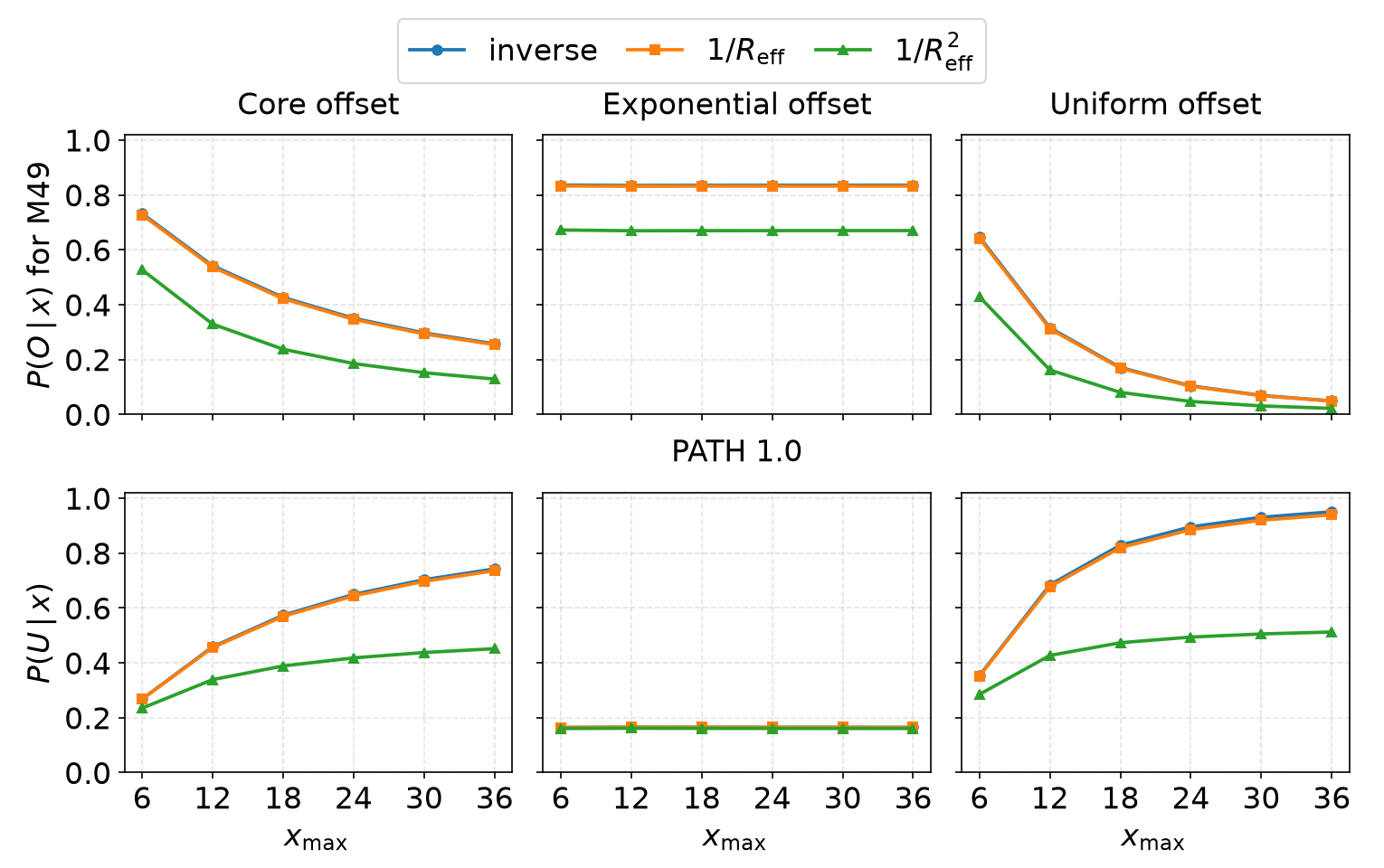}
    \hfill % Pushes the images to opposite sides, acting as a spacer
    % Figure 1b (Right) - No blank lines between this and the previous image
    \includegraphics[width=0.45\textwidth]{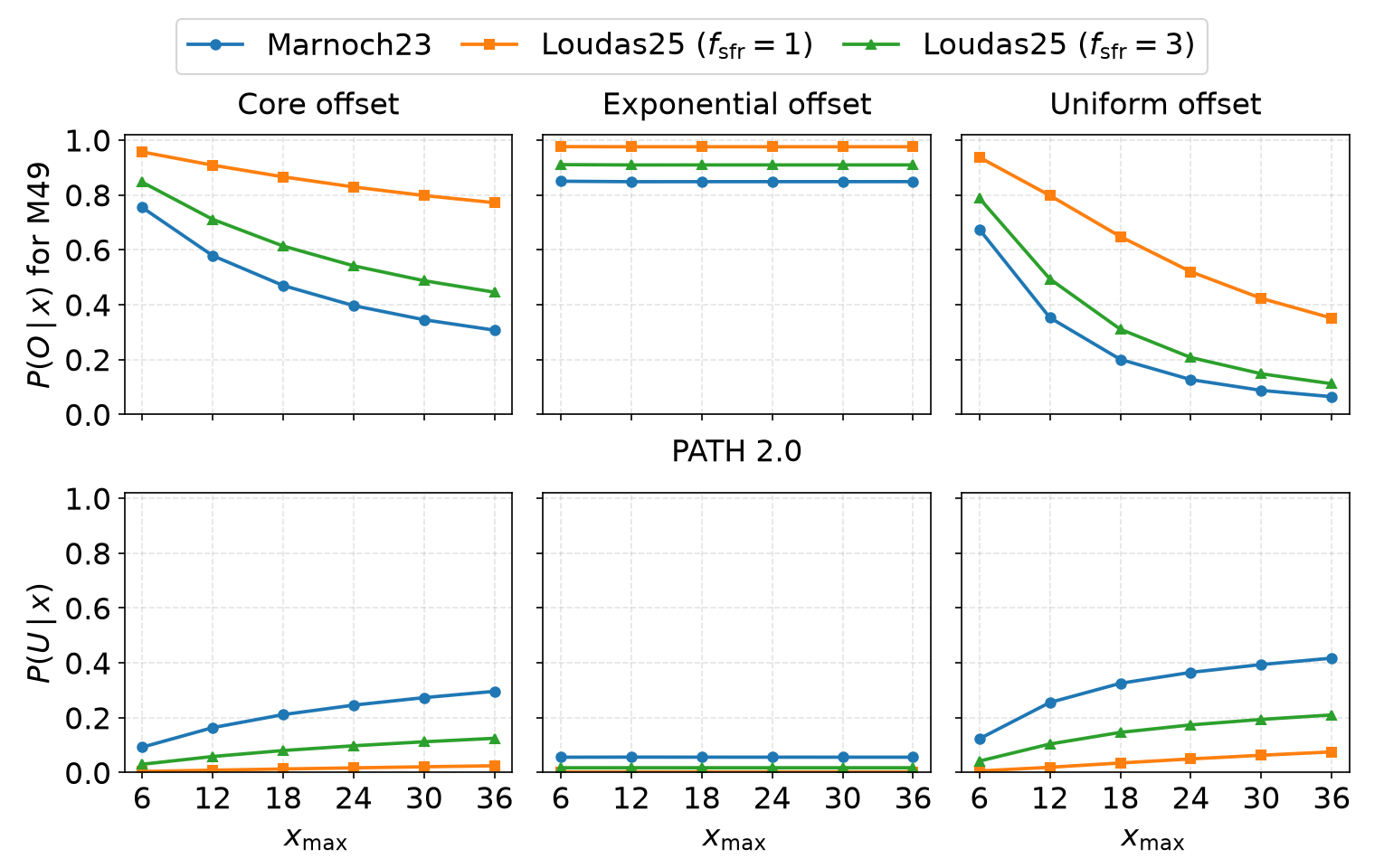}
    
    \caption{
        \textbf{(a)} \textit{Left:} Posterior probabilities for M49 across all evaluated PATH 1.0 prior configurations, demonstrating that M49 remains the preferred candidate host for nearly all physically motivated prior configurations. 
        \textbf{(b)} \textit{Right:} Posterior probabilities for M49 across PATH 2.0 prior configurations, demonstrating that PATH 2.0 similarly favors M49 for nearly all physically motivated prior configurations.
    }
    \label{fig:path_analysis}
\end{figure*}

\begin{figure}
\centering
\includegraphics[width = 0.462\textwidth]{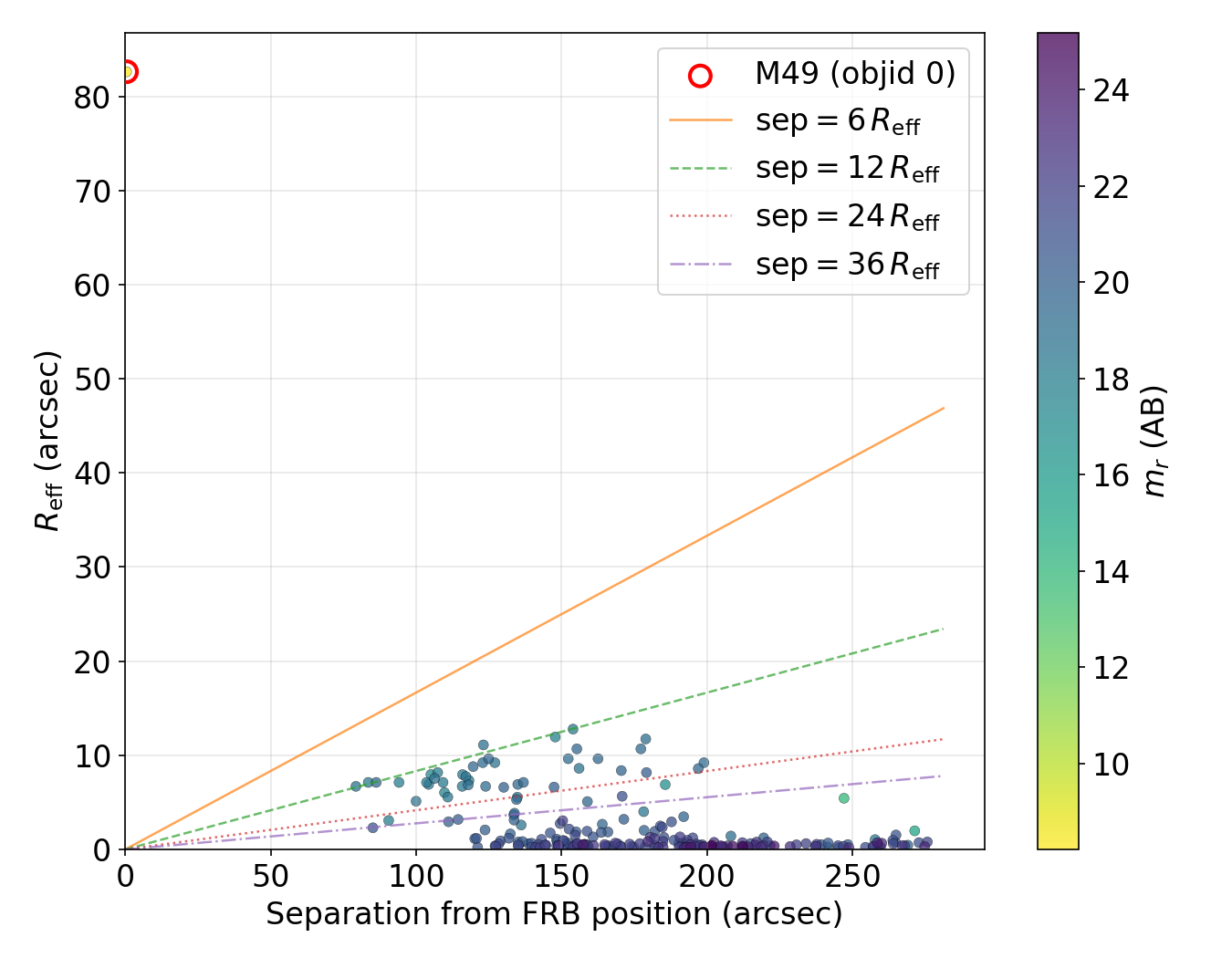}
\caption{Distribution of candidate galaxies in effective radius versus separation. Lines indicate the truncation limits from the localisation centre. Only a minority of galaxies possess truncation radii large enough to encompass the localisation centre.}
\label{candidate_re_vs_sep}
\end{figure}

The PATH framework \citep{Aggarwal2021} evaluates hosts by combining the candidate prior $P(O_i)$, the offset prior $P(x|O_i)$, and an unseen prior $P(U)$. PATH integrates over the FRB error localisation region to determine the posteriors ($P(O_i|x)$ and $P(U|x)$). The candidate prior assigns higher baseline probabilities to brighter galaxies, scaling inversely with the surface density of galaxies brighter than the candidate ($P(O_i) \propto 1/\Sigma_m$). Optionally, the candidate prior can incorporate an inverse dependence on $R_{\mathrm{eff}}$, further scaling as $1/R_{\mathrm{eff}}$ or $1/R_{\mathrm{eff}}^2$. PATH offers three offset priors, each evaluated as a function of the absolute separation, $\theta$. The separation is normalised by $R_{\mathrm{eff}}$ to give $x = \theta/R_{\mathrm{eff}}$, with integration truncated at a maximum separation of $\theta_{\mathrm{max}} = x_{\mathrm{max}} \times R_{\mathrm{eff}}$. The Exponential model scales with $\mathrm{e}^{-x}$, the Core model scales with $(1 + x)^{-1}$, while the Uniform model assigns equal probabilities up to truncation. We run PATH on all combinations of these priors with $P(U) = 0.2$, as detailed in Section~\ref{sec:search} using a top-hat localisation distribution. As shown in Figure~\ref{fig:path_analysis}a, M49 remains the preferred host for nearly all prior configurations.

To investigate this preference, we first isolate the geometry-independent candidate prior, $P(O_i)$. Due to M49's high brightness, it maintains a candidate prior greater than $0.99$ post-normalisation under both the standard inverse prior and further scaling by $1/R_{\mathrm{eff}}$. Even under the more extreme $1/R_{\mathrm{eff}}^2$ scaling, it retains a prior greater than $0.8$. This demonstrates that even before considering geometry, PATH highly favours M49 as the host. We note that for large localisation regions (such as 4.6\arcmin), $P(O|x)$ becomes dominated by the prior distribution \citep{Hewitt-Bhardwaj}. We next incorporate the localisation distribution by studying the offset priors, $P(x|O_i)$. All galaxies in the field other than M49 have large normalised separations ($x_i \gtrsim 10$) from the FRB signal centre, as shown in Figure~\ref{candidate_re_vs_sep}. The $m_r$ colour scale in the plot helps clarify both the candidate prior and the field geometry. Under the Exponential model, these galaxies are heavily suppressed and effectively removed from consideration. This causes M49 to maintain a posterior greater than $0.8$ independent of $x_\mathrm{max}$. However, the Core and Uniform models apply a much shallower penalty on $x_i$. M49's large angular size ($R_{\mathrm{eff}} \approx 82\arcsec$) allows its spatial prior to completely encompass the FRB localisation well within the PATH default $x_\mathrm{max} = 6$. Increasing the truncation limit further simply extends this prior over a larger integration area, diluting M49's local probability density. While expanding $x_\mathrm{max}$ also increases the number of statistically viable candidates, the dilution of M49's large share of the probability density causes the total marginal likelihood, $p(x)$, to decrease. Ultimately, increasing the truncation limit redistributes M49's share of the probability between the unseen posterior ($P(U|x)$) and other viable candidates. Even accounting for this dilution, M49 remains the most preferred host under $52$ of $54$ configuration runs. It is only outranked by another host candidate with a margin of less than $2.5$ per cent under the most aggressive combination of penalties: the Uniform model coupled with $1/R_\mathrm{eff}^2$ scaling at $x_\mathrm{max} = 30$ and $36$. 

We recognise that integrating to arbitrarily large truncation limits is physically unmotivated. As demonstrated by FRB 20200120E \citep{Kirsten2022}, progenitors do not strictly follow the primary stellar light distribution and can originate in extended environments. To establish an upper limit for the separation of bound galactic matter, we explore GC systems, as they are observed to be substantially more extended than the main stellar body. Defining $R_{\mathrm{eff, GCS}}$ as the projected half-number radius of a host galaxy's GC system, observations indicate $R_{\mathrm{eff, GCS}} \approx 3\text{--}5 R_{\mathrm{eff}}$ \citep{Forbes2017}. We therefore consider $x_\mathrm{max} = 12$ to be a physically motivated upper limit for the truncation radius, corresponding to approximately $2-4\,R_\mathrm{eff, GCS}$ and encompassing the majority of the expected GC population. The adopted limit is a deliberately conservative upper bound when compared to the empirical FRB population, where the median FRB offset is $2.09 R_{\mathrm{eff}}$ \citep{Woodland2024}, and the largest offset measured to date is $5.7 R_{\mathrm{eff}}$ \citep{Shah2025}. 

When restricted to this physically motivated region ($x_\mathrm{max}  \leq 12$), M49 remains the top host candidate for all prior configurations and maintains a posterior greater than $P(U|x)$ for a majority of runs. We repeat our analysis using the new PATH 2.0 framework \citep{James_Loudas_Woodland_Andersen_Prochaska_Hoffmann_Marnoch_Ryder_2026}, and find that M49 remains the top host candidate, and is assigned a posterior greater than that from PATH 1.0 for a majority of runs, as shown in Figure~\ref{fig:path_analysis}b. We note that the Naive prior configuration assigns a candidate prior of exactly 0 to M49, thereby removing it from further Bayesian analysis and constraining the posteriors to 0. The Naive model assigns $P(O_i) = 0$ to all candidates with $M_r < -23$, and we therefore omit it from the plot of posteriors. 

This case illustrates a limitation of PATH when applied to fields containing massive foreground galaxies. The large angular sizes and high brightness of these galaxies skew the spatial and candidate priors to such an extent that they suppress all other candidates even with large localisation errors. We advise caution when relying solely on PATH for host association in such environments.

\section{Galaxies in the Field of FRB~20240817A}\label{app:lowsiggals}

\begin{table}
    \centering
    \begin{tabular}{lccccc} 
 {ID} & {RA} & {DEC} & {Magnitude} & {Angular Size} & {Separation} \\ 
  & {degrees} & {degrees} & AB Band & arcsec$^2$ & arcsec \\
 \hline
0 & 187.4449 & 8.0005 & 8.22 & 82.61 & 0.49 \\
1 & 187.3760 & 7.9928 & 13.92 & 5.44 & 247.04 \\
2 & 187.3958 & 8.058 &15.22 & 1.98 & 271.34 \\
3 & 187.4133 & 8.0414 & 16.83 & 6.89 & 185.62 \\
4 & 187.4170 & 7.9909 & 17.55 & 7.96 & 105.09 \\
5 & 187.4425 & 7.9700 & 17.86 & 6.05 & 109.76 \\
6 & 187.4566 & 7.961 & 17.87 & 11.95 & 147.82 \\
7 & 187.4251 & 7.9847 & 17.99 & 3.08 & 90.56 \\
8 & 187.4290 & 7.9795 & 18.06 & 7.13 & 94.13 \\
9 & 187.4253 & 8.0231 & 18.19 & 8.17 & 107.41 \\
10 & 187.474 & 7.9688 & 18.27 & 12.79 & 153.92 \\
11 & 187.4459 & 7.9682 & 18.31 & 7.96 & 115.96 \\
12 & 187.4523 & 7.971 & 18.34 & 7.13 & 109.25 \\
13 & 187.4494 & 7.9718 & 18.34 & 6.92 & 104.31 \\
14 & 187.4334 & 8.0268 & 18.41 & 7.13 & 103.58 \\
15 & 187.4954 & 8.0516 & 18.43 & 1.05 & 257.70 \\
16 & 187.4286 & 7.9838 & 18.45 & 7.13 & 83.56 \\
17 & 187.4178 & 7.9852 & 18.48 & 5.57 & 110.89 \\
18 & 187.4235 & 8.0254 & 18.49 & 7.34 & 118.16 \\
19 & 187.3981 & 8.0178 & 18.5 & 4.021 & 178.20 \\
20 & 187.4318 & 7.9709 & 18.53 & 6.71 & 115.84 \\
21 & 187.4117 & 8.0190 & 18.68 & 2.6 & 136.10 \\
22 & 187.4104 & 7.9915 & 18.69 & 9.22 & 127.05 \\ 
23 & 187.4676 & 7.9841 & 18.70 & 5.13 & 99.98 \\ 
24 & 187.4669 & 7.9969 & 18.71 & 6.71 & 79.32 \\ 
25 & 187.4248 & 8.0321 & 18.73 & 5.55 & 134.79 \\ 
26 & 187.4611 & 7.9602 & 18.78 & 8.62 & 155.99 \\ 
27 & 187.4743 & 8.0054 & 18.95 & 7.55 & 106.33 \\ 
28 & 187.4146 & 7.9878 & 18.97 & 7.74 & 117.11 \\ 
29 & 187.4104 & 8.0008 & 19.03 & 11.11 & 123.13 \\ 
30 & 187.4704 & 7.9576 & 19.09 & 11.74 & 178.91 \\ 
31 & 187.4768 & 7.9917 & 19.09 & 6.86 & 117.93 \\ 
32 & 187.4771 & 7.9522 & 19.19 & 1.43 & 208.17 \\ 
33 & 187.4502 & 7.9584 & 19.3 & 9.64 & 152.27 \\ 
34 & 187.4394 & 7.9664 & 19.33 & 6.71 & 123.89 \\ 
35 & 187.4164 & 7.9812 & 19.34 & 9.22 & 122.91 \\ 
36 & 187.4821 & 7.9592 & 19.38 & 9.22 & 198.91 \\ 
37 & 187.4809 & 7.9664 & 19.40 & 10.69 & 177.23 \\ 
38 & 187.4700 & 7.9729 & 19.44 & 3.68 & 133.55 \\ 
39 & 187.4648 & 7.9685 & 19.47 & 6.92 & 134.94 \\ 
40 & 187.4147 & 8.0228 & 19.51 & 5.31 & 134.48 \\ 
41 & 187.4765 & 8.0114 & 19.53 & 8.80 & 119.6 \\ 
42 & 187.3995 & 8.0075 & 19.55 & 2.68 & 163.98 \\ 
43 & 187.4100 & 7.9969 & 19.56 & 9.64 & 124.91 \\ 
44 & 187.4799 & 7.9715 & 19.56 & 9.64 & 162.52 \\ 
46 & 187.4674 & 7.9623 & 19.56 & 5.09 & 158.85 \\ 
47 & 187.479 & 7.9573 & 19.61 & 8.59 & 196.96 \\ 
48 & 187.4956 & 7.9669 & 19.61 & 0.24 & 217.35 \\ 
49 & 187.4287 & 8.0404 & 19.62 & 10.69 & 155.24 \\ 
50 & 187.3716 & 7.9829 & 19.69 & 0.46 & 268.97 \\ 
51 & 187.4664 & 7.9689 & 19.70 & 7.13 & 136.88 \\ 
52 & 187.4376 & 8.0369 & 19.71 & 3.85 & 133.81 \\ 
53 & 187.4760 & 7.9569 & 19.73 & 3.48 & 191.94 \\ 
54 & 187.5116 & 7.9886 & 19.76 & 0.26 & 241.36 \\ 
55 & 187.4603 & 7.9556 & 19.79 & 8.38 & 170.5 \\ 
56 & 187.4876 & 8.0443 & 19.83 & 1.24 & 219.42 \\ 
57 & 187.476 & 8.0027 & 19.83 & 2.95 & 111.14 \\ 
58 & 187.4743 & 7.979 & 19.90 & 6.60 & 130.11 \\ 
59 & 187.4828 & 7.9987 & 19.92 & 0.80 & 135.08 \\ 
60 & 187.4914 & 7.9822 & 19.95 & 2.04 & 178.34 \\ 
61 & 187.4815 & 7.9943 & 19.97 & 1.66 & 132.38 \\ 
62 & 187.4075 & 7.998 & 20.05 & 3.13 & 133.67 \\ 
63 & 187.4590 & 8.0318 & 20.06 & 2.08 & 123.68 \\

\end{tabular}
    \caption{From left to right, we list the galaxy ID, right ascension, declination, AB magnitude, angular size, and the separation between the centre of the galaxy and FRB~20240817A.}
    \label{tab:galaxyIDs}
\end{table}

\begin{table}
    \centering
    \begin{tabular}{lccccc} 
 {ID} & {RA} & {DEC} & {Magnitude} & {Angular Size} & {Separation} \\ 
  & {degrees} & {degrees} & AB Band & arcsec$^2$ & arcsec \\
 \hline
64 & 187.4175 & 8.0311 & 20.08 & 6.63 & 147.4 \\
65 & 187.4334 & 8.0466 & 20.15 & 3.21 & 171.38 \\
66 & 187.4674 & 8.0231 & 20.19 & 3.19 & 114.43 \\ 
67 & 187.4822 & 7.9670 & 20.21 & 8.17 & 179.15 \\ 
68 & 187.4266 & 8.067 & 20.32 & 0.53 & 248.38 \\ 
69 & 187.4747 & 7.9848 & 20.34 & 1.15 & 120.16 \\ 
70 & 187.4418 & 7.9539 & 20.35 & 0.31 & 167.7 \\ 
71 & 187.3772 & 7.9745 & 20.41 & 0.43 & 258.81 \\ 
72 & 187.4408 & 7.9621 & 20.44 & 0.63 & 138.65 \\ 
73 & 187.4759 & 7.9557 & 20.48 & 0.26 & 195.15 \\ 
74 & 187.4657 & 7.9626 & 20.49 & 1.88 & 154.95 \\ 
75 & 187.4101 & 8.0310 & 20.49 & 1.86 & 165.91 \\ 
76 & 187.4694 & 7.9671 & 20.53 & 1.10 & 148.23 \\ 
77 & 187.4091 & 7.987 & 20.54 & 0.84 & 136.56 \\ 
78 & 187.5103 & 7.9796 & 20.55 & 0.2 & 244.99 \\ 
79 & 187.3783 & 8.0082 & 20.57 & 0.52 & 239.21 \\ 
80 & 187.4508 & 7.9775 & 20.57 & 2.31 & 85.16 \\ 
81 & 187.4507 & 7.9335 & 20.58 & 0.69 & 241.58 \\ 
82 & 187.4091 & 8.0289 & 20.60 & 1.81 & 163.57 \\ 
83 & 187.4267 & 8.0385 & 20.61 & 0.56 & 151.73 \\ 
84 & 187.4732 & 7.9780 & 20.64 & 0.92 & 129.01 \\ 
85 & 187.4546 & 8.0327 & 20.67 & 0.24 & 121.15 \\ 
86 & 187.4489 & 7.9559 & 20.69 & 1.34 & 160.80 \\ 
87 & 187.4865 & 7.9987 & 20.7 & 0.85 & 148.42 \\ 
88 & 187.3977 & 7.9951 & 20.71 & 0.6 & 169.26 \\ 
89 & 187.4755 & 8.0148 & 20.72 & 1.19 & 120.77 \\ 
90 & 187.4835 & 7.9674 & 20.73 & 1.93 & 182.02 \\ 
91 & 187.4077 & 7.9882 & 20.74 & 0.66 & 139.73 \\ 
92 & 187.4481 & 7.9586 & 20.76 & 0.58 & 151.04 \\ 
93 & 187.4236 & 7.9519 & 20.76 & 0.66 & 190.24 \\ 
94 & 187.4700 & 7.9683 & 20.77 & 1.76 & 146.11 \\ 
95 & 187.3757 & 8.0271 & 20.84 & 0.75 & 264.73 \\ 
96 & 187.4702 & 7.9669 & 20.84 & 0.93 & 150.84 \\ 
97 & 187.4019 & 7.9404 & 20.85 & 1.54 & 264.89 \\ 
98 & 187.4129 & 8.0289 & 20.86 & 0.21 & 153.60 \\ 
99 & 187.4815 & 8.0056 & 20.86 & 1.20 & 131.84 \\ 
100 & 187.4173 & 8.0351 & 20.87 & 0.24 & 159.18 \\ 
101 & 187.3938 & 8.0426 & 20.87 & 0.29 & 237.41 \\ 
102 & 187.4831 & 7.9631 & 20.9 & 0.66 & 191.45 \\ 
103 & 187.3934 & 7.9478 & 20.90 & 0.97 & 263.81 \\ 
104 & 187.4277 & 7.96 & 20.95 & 0.54 & 158.02 \\ 
105 & 187.4207 & 8.047 & 20.97 & 1 & 188.66 \\ 
106 & 187.4017 & 8.0111 & 20.98 & 1.94 & 158.73 \\ 
107 & 187.4932 & 7.9731 & 21 & 0.51 & 198.32 \\ 
108 & 187.4462 & 7.9324 & 21.02 & 0.41 & 244.97 \\ 
109 & 187.4056 & 8.0172 & 21.02 & 0.66 & 152.52 \\ 
110 & 187.3968 & 8.0091 & 21.03 & 0.34 & 174.18 \\ 
111 & 187.3736 & 7.9965 & 21.04 & 0.24 & 254.41 \\ 
112 & 187.4041 & 8.0253 & 21.05 & 5.66 & 170.80 \\ 
113 & 187.4062 & 8.0156 & 21.05 & 0.4 & 148.34 \\ 
114 & 187.4724 & 7.9714 & 21.06 & 1.07 & 143.08 \\ 
115 & 187.4030 & 7.9981 & 21.08 & 2.76 & 149.49 \\ 
116 & 187.4710 & 7.9494 & 21.09 & 0.22 & 205.94 \\ 
117 & 187.4697 & 7.9527 & 21.12 & 0.68 & 193.15 \\ 
118 & 187.4868 & 7.9706 & 21.12 & 2.50 & 183.94 \\ 
119 & 187.4683 & 7.9537 & 21.13 & 2.94 & 187.76 \\ 
120 & 187.4838 & 7.9828 & 21.14 & 2.17 & 152.54 \\ 
121 & 187.4258 & 7.9706 & 21.17 & 0.39 & 127.13 \\ 
122 & 187.4480 & 7.9576 & 21.18 & 0.43 & 154.38 \\ 
123 & 187.4585 & 7.9258 & 21.19 & 0.73 & 272.78 \\ 
124 & 187.4104 & 7.9770 & 21.22 & 0.37 & 148.85 \\ 
125 & 187.4751 & 7.9495 & 21.24 & 0.26 & 212.63 \\ 
126 & 187.4525 & 7.9549 & 21.24 & 0.40 & 165.98 \\ 
127 & 187.4347 & 8.054 & 21.24 & 0.39 & 196.27 \\ 
 
\end{tabular}
    \caption{-- {\it continued}}
\end{table}

\begin{table}
    \centering
    \begin{tabular}{lccccc} 
 {ID} & {RA} & {DEC} & {Magnitude} & {Angular Size} & {Separation} \\ 
  & {degrees} & {degrees} & AB Band & arcsec$^2$ & arcsec \\
 \hline
128 & 187.4949 & 7.9866 & 21.25 & 1.29 & 185.17 \\
129 & 187.4138 & 7.9789 & 21.25 & 0.36 & 135.14 \\ 
130 & 187.5047 & 8.0081 & 21.25 & 0.63 & 214.90 \\ 
131 & 187.4210 & 8.0377 & 21.26 & 0.44 & 158.95 \\ 
132 & 187.3944 & 8.0116 & 21.26 & 2.36 & 184.58 \\ 
133 & 187.4849 & 8.0320 & 21.28 & 0.25 & 182.32 \\ 
134 & 187.4061 & 8.0057 & 21.31 & 0.25 & 139.70 \\ 
135 & 187.4384 & 7.9503 & 21.32 & 0.44 & 181.94 \\ 
136 & 187.4351 & 8.0731 & 21.32 & 0.63 & 263.93 \\ 
137 & 187.4793 & 7.9898 & 21.33 & 0.6 & 128.43 \\ 
138 & 187.4828 & 8.001 & 21.34 & 0.48 & 135.1 \\ 
139 & 187.4011 & 8.0065 & 21.35 & 0.55 & 157.75 \\ 
140 & 187.4863 & 7.9875 & 21.37 & 1.52 & 154.65 \\ 
141 & 187.4029 & 8.0119 & 21.38 & 0.37 & 155.42 \\ 
142 & 187.4658 & 7.9615 & 21.39 & 0.47 & 158.58 \\ 
143 & 187.3987 & 8.0030 & 21.40 & 0.34 & 164.95 \\ 
144 & 187.4862 & 7.9649 & 21.41 & 1.24 & 195.07 \\ 
145 & 187.4813 & 7.9843 & 21.42 & 0.53 & 142.08 \\ 
146 & 187.4428 & 8.0421 & 21.42 & 0.95 & 150.41 \\ 
147 & 187.4717 & 7.9501 & 21.43 & 0.24 & 204.78 \\ 
148 & 187.5017 & 7.9754 & 21.49 & 0.41 & 221.65 \\ 
149 & 187.4093 & 8.0035 & 21.49 & 0.34 & 127.4 \\ 
150 & 187.4837 & 7.9887 & 21.51 & 0.47 & 144.58 \\ 
151 & 187.4804 & 7.9502 & 21.54 & 0.81 & 220.70 \\ 
152 & 187.4055 & 8.0311 & 21.60 & 0.25 & 178.64 \\ 
153 & 187.3971 & 7.9991 & 21.61 & 0.30 & 170.41 \\ 
154 & 187.4054 & 8.0027 & 21.63 & 0.21 & 141.1 \\ 
155 & 187.4694 & 7.9363 & 21.70 & 0.34 & 246.86 \\ 
156 & 187.4539 & 7.9527 & 21.74 & 0.22 & 174.85 \\ 
157 & 187.3947 & 7.9457 & 21.75 & 0.4 & 267.89 \\ 
158 & 187.4720 & 7.948 & 21.8 & 0.31 & 212.04 \\ 
159 & 187.4621 & 8.0751 & 21.80 & 0.81 & 275.68 \\ 
160 & 187.4895 & 7.9833 & 21.81 & 0.34 & 170.31 \\ 
161 & 187.4065 & 8.0315 & 21.83 & 0.63 & 176.92 \\ 
162 & 187.3959 & 8.0423 & 21.85 & 0.50 & 230.70 \\ 
163 & 187.4249 & 7.9318 & 21.85 & 0.28 & 257.07 \\ 
164 & 187.4872 & 7.9892 & 21.86 & 0.21 & 156.11 \\ 
165 & 187.5080 & 8.0182 & 21.87 & 0.63 & 234 \\ 
166 & 187.3947 & 8.0206 & 21.88 & 0.98 & 193.16 \\ 
167 & 187.4943 & 7.9764 & 21.9 & 0.58 & 196.36 \\ 
168 & 187.4320 & 8.0468 & 21.90 & 0.57 & 173.37 \\ 
169 & 187.4137 & 7.9756 & 21.91 & 0.18 & 142.53 \\ 
170 & 187.4049 & 7.994 & 21.94 & 0.46 & 144.33 \\ 
171 & 187.4143 & 8.0590 & 21.96 & 0.55 & 237.6 \\ 
172 & 187.3951 & 8.0095 & 21.97 & 1.30 & 180.45 \\ 
173 & 187.4936 & 7.9829 & 21.99 & 0.85 & 184.82 \\ 
174 & 187.4848 & 7.9409 & 22.00 & 0.34 & 257.07 \\ 
175 & 187.5036 & 7.9909 & 22.01 & 0.73 & 212.10 \\ 
176 & 187.4749 & 7.9710 & 22.04 & 3.06 & 150.37 \\ 
177 & 187.4456 & 8.0435 & 22.04 & 0.57 & 155.24 \\ 
178 & 187.4948 & 7.9813 & 22.07 & 1.34 & 190.7 \\ 
179 & 187.4332 & 8.0527 & 22.09 & 0.27 & 192.86 \\ 
180 & 187.3905 & 8.0109 & 22.17 & 0.42 & 197.49 \\ 
181 & 187.4019 & 8.0039 & 22.21 & 0.29 & 153.95 \\ 
182 & 187.4212 & 8.0512 & 22.21 & 0.31 & 201.30 \\ 
183 & 187.3991 & 8.0062 & 22.22 & 0.53 & 164.76 \\ 
184 & 187.5026 & 8.0068 & 22.24 & 0.39 & 206.97 \\ 
185 & 187.3922 & 8.0089 & 22.25 & 0.24 & 190.41 \\ 
186 & 187.3956 & 8.0135 & 22.26 & 1.16 & 181.91 \\ 
187 & 187.4911 & 7.9602 & 22.28 & 0.41 & 219.26 \\ 
188 & 187.4015 & 7.9958 & 22.29 & 0.64 & 155.45 \\ 
189 & 187.4003 & 7.9850 & 22.31 & 0.68 & 168.42 \\ 
190 & 187.4701 & 8.0375 & 22.33 & 0.42 & 161 \\ 
191 & 187.4384 & 7.9464 & 22.35 & 0.34 & 195.72 \\ 
192 & 187.4817 & 7.9610 & 22.37 & 0.28 & 193.23 \\ 
 
\end{tabular}
    \caption{-- {\it continued}}
\end{table}

\begin{table}
    \centering
    \begin{tabular}{lccccc} 
 {ID} & {RA} & {DEC} & {Magnitude} & {Angular Size} & {Separation} \\ 
  & {degrees} & {degrees} & AB Band & arcsec$^2$ & arcsec \\
 \hline
193 & 187.3868 & 7.9826 & 22.37 & 0.79 & 216.98 \\ 
194 & 187.3938 & 8.0089 & 22.40 & 0.25 & 184.87 \\
195 & 187.3917 & 8.0119 & 22.41 & 0.51 & 194.23 \\ 
196 & 187.4775 & 7.9429 & 22.41 & 0.35 & 237.31 \\ 
197 & 187.3789 & 7.9653 & 22.46 & 0.41 & 267.02 \\ 
198 & 187.5173 & 8.0075 & 22.50 & 0.24 & 259.55 \\ 
199 & 187.49 & 7.968 & 22.58 & 0.27 & 198.71 \\ 
200 & 187.4239 & 7.9646 & 22.72 & 0.40 & 148.97 \\ 
201 & 187.4709 & 8.0584 & 22.74 & 0.32 & 228.49 \\ 
202 & 187.3936 & 8.0041 & 22.76 & 0.21 & 183.22 \\ 
203 & 187.3953 & 8.0277 & 22.83 & 0.76 & 202.4 \\ 
204 & 187.4188 & 8.0471 & 22.83 & 0.17 & 192.15 \\ 
205 & 187.3922 & 8.0302 & 22.89 & 0.47 & 216.53 \\ 
206 & 187.4000 & 7.9777 & 22.91 & 0.84 & 179.62 \\ 
207 & 187.5013 & 7.9924 & 22.96 & 0.18 & 203.17 \\ 
208 & 187.4959 & 8.0521 & 22.96 & 0.66 & 260.02 \\ 
209 & 187.3881 & 8.0094 & 23.06 & 0.32 & 205.16 \\ 
210 & 187.4755 & 7.9478 & 23.11 & 0.27 & 218.62 \\ 
211 & 187.3921 & 7.9552 & 23.12 & 0.18 & 248.77 \\ 
212 & 187.5027 & 7.95 & 23.26 & 0.30 & 274.76 \\ 
213 & 187.388 & 8.0079 & 23.27 & 0.4 & 204.71 \\ 
214 & 187.4980 & 7.9903 & 23.42 & 0.22 & 192.78 \\ 
215 & 187.42 & 8.0545 & 23.56 & 0.28 & 214.07 \\ 
216 & 187.4194 & 8.0496 & 23.57 & 0.29 & 199.09 \\ 
217 & 187.4968 & 7.9503 & 23.75 & 0.73 & 258.49 \\ 
218 & 187.5028 & 7.9768 & 23.90 & 0.37 & 223.08 \\ 
219 & 187.4205 & 8.0369 & 24.05 & 0.51 & 157.39 \\ 
220 & 187.3909 & 8.008 & 24.06 & 0.23 & 194.42 \\ 
221 & 187.3864 & 8.0043 & 24.47 & 0.36 & 208.84 \\ 
222 & 187.5032 & 7.9885 & 24.87 & 0.36 & 212.21 \\ 
223 & 187.3886 & 8.0056 & 25.18 & 0.44 & 201.52 \\

\end{tabular}
    \caption{-- {\it continued}}
\end{table}

% Don't change these lines
\bsp    % typesetting comment
\label{lastpage}
\end{document}